\documentclass[11pt]{article}

\usepackage[a4paper,margin=25mm]{geometry}
\usepackage[T1]{fontenc}
\usepackage[utf8]{inputenc}
\usepackage{microtype}
\usepackage{booktabs}
\usepackage{graphicx}
\usepackage{xcolor}
\usepackage{xspace}
\usepackage{amsmath}
\usepackage{multirow}
\usepackage{makecell}
\usepackage{array}
\usepackage[numbers,sort&compress]{natbib}
\usepackage{tikz}
\usetikzlibrary{patterns,arrows.meta,decorations.pathreplacing}
\PassOptionsToPackage{hyphens}{url}
\usepackage[hidelinks]{hyperref}

\definecolor{obOrange}{HTML}{E69F00}
\definecolor{obBlue}{HTML}{56B4E9}
\definecolor{obGreen}{HTML}{009E73}
\definecolor{obDark}{HTML}{D55E00}

\newcommand{\ucache}{uCache\xspace}
\newcommand{\xcache}{XCache\xspace}
\newcommand{\datasetZ}{AGC\,2015\xspace}      
\newcommand{\datasetL}{UL2016\xspace}         
\newcommand{\mD}{D\xspace}  
\newcommand{\mW}{W\xspace}  
\newcommand{\mFone}{F$_1$\xspace}  
\newcommand{\mFtwo}{F$_2$\xspace}  
\newcommand{\mV}{V\xspace}  
\newcommand{\mM}{M\xspace}  
\newcommand{\z}{\phantom{0}}

\hypersetup{
  pdftitle={Client-side transparent caching for remote ROOT data analysis},
  pdfauthor={Dmytro Kovalskyi, Jan Eysermans, Mariarosaria D'Alfonso,
    Christoph Paus}}

\title{Client-side transparent caching for remote ROOT data analysis}

\author{Dmytro Kovalskyi$^{1}$\thanks{Corresponding author: \texttt{kdv@mit.edu}},
  Jan Eysermans$^{1,2}$, Mariarosaria D'Alfonso$^{1}$, Christoph Paus$^{1}$\\[4pt]
  \small $^{1}$Massachusetts Institute of Technology, Cambridge, MA, USA\\
  \small $^{2}$CERN, Geneva, Switzerland}

\date{}

\begin{document}
\maketitle

\begin{abstract}
  \noindent
High-energy physics analyses often process the same data as physicists refine
algorithms and test new ideas.  With data increasingly read from remote storage,
each iteration is subject to network latency and depends on network bandwidth
and shared-storage throughput, which can vary substantially under load. We
present \ucache (\texttt{xrd-ucache}), a transparent client-side cache
implemented as an XRootD client plugin that requires neither server-side
deployment nor changes to analysis code. It uses local storage on the analysis
machine as a cache layer between the network and memory. The cache stores only
the data actually read by an analysis. It can also rebuild cached data into a
branch-aligned, recompressed form, eliminating most of the input/output and
decompression costs of subsequent passes. We benchmark the cache using the
Analysis Grand Challenge top quark pair analysis on public CMS Open Data
compressed with zlib and LZMA. Filling the cache adds essentially no overhead
compared with a direct read. Subsequent passes are 1.6--8.8 times faster from
the byte cache and 2.1--15.7 times faster from the recompressed cache. For a
typical analysis, a 1\,TB cache suffices for datasets of 10--20\,TB. The largest
improvements occur when the remote data source is heavily loaded or
geographically distant.
\end{abstract}

\section{Introduction}
\label{sec:intro}

High-energy physics analysis is an iterative process. A result is rarely
produced by writing the final analysis once and running it over the
data. Instead, physicists develop an idea, implement it in code, process the data,
inspect the output and use what they learn to decide what to try next. Each
cycle may involve changing an event selection, testing a different observable,
comparing alternative background models, checking a possible bias or
investigating an unexpected feature in the data.

The speed of this cycle directly affects how thoroughly an analysis can be
explored. When each iteration takes a long time, the number of ideas that can be
tested is limited; when feedback is fast, physicists can explore more
alternatives, investigate unexpected observations, and identify weak approaches
or possible biases earlier. Faster feedback allows physicists to test more
alternatives and make better-informed choices in the final analysis.

Physicists working on experiments at the CERN LHC analyze many
billions of events. Even in a compact format such as CMS
NanoAOD~\cite{rizzi2019nanoaod}, the data used by a single analysis can total
tens of terabytes, reaching the 100\,TB scale. Processing such large
volumes typically relies on local computing clusters or distributed computing
resources, and producing complete results can take considerable time. Additional
delays can arise from transient failures or periods of high demand for computing
resources.

Resource contention is particularly important because shared computing systems
are expensive. Such systems are generally optimized for scheduled, large-scale
processing rather than on-demand tasks that require rapid turnaround. Even a
relatively small or urgent request may have to wait until sufficient resources
become available. Interactive data analysis is therefore often performed on
personal laptops or dedicated workstations, where users have direct control over
the available resources and can prioritize workloads according to their
immediate needs.

To achieve acceptable turnaround, physicists use several methods to reduce the
amount of data to process. The two most important methods are event skimming and
slimming. Event skimming selects a small subset of relevant events by applying
preselection requirements, while slimming reduces the size of each event by
removing information that is not needed for the analysis.

Both approaches can be highly effective. However, reducing a dataset to a size
that can be processed on an individual computer often requires a reduction of
3 to 4 orders of magnitude. Such a large reduction can restrict
experimentation because less information remains available for testing
alternative analysis strategies. It also commonly requires an additional,
time-consuming processing step to produce so-called final flat ntuples, which
are then used for statistical inference and other final analysis tasks.

To mitigate these issues, column-based ntuples are widely used. These ntuples
store data by branch, allowing rapid access to only the required
information. Although columnar ntuples such as NanoAOD can be read very
efficiently~\cite{brun1997root,piparo2019rdataframe}, they require a significant
amount of storage to retain all information that may be useful. For a
collaboration-wide data format such as NanoAOD, this often means that only a
small fraction of the stored data are needed for a particular analysis, despite
the large amount of disk space required for the complete dataset.

These datasets can be read directly from remote storage using the XRootD
protocol~\cite{dorigo2005xrootd}, which avoids the need to keep a full local
copy. However, each pass over the data then depends on the network and the remote
storage system: file opens alone can add tens of seconds per pass over a
wide-area link (Section~\ref{sec:openlatency}), the throughput of a shared
origin can change by a factor of a few within an hour
(Section~\ref{sec:origins}), and a temporary server error can interrupt an
entire job.
A local cache reduces these problems by retrying failed
reads and serving repeatedly accessed data locally, without contacting the
remote server again.

The case for placing that cache on the analysis machine itself is
structural. A modern NVMe device delivers several gigabytes per second at
tens of microseconds: more than a 10\,GbE network path carries, and far
more than a fair per-client share of any shared network layer in a balanced
deployment (Section~\ref{sec:xcache} measures the network case directly).
Local storage is therefore the fastest layer an physicist can place between
the machine's memory and the remote infrastructure. Today that layer
usually serves as scratch space; \ucache turns it into a durable cache.

Analysts can use \ucache to turn local SSD or NVMe storage into a fast cache
on a laptop or workstation. It is implemented as an XRootD client plugin that end
users can install without administrator support, server-side modifications or
changes to the analysis code. Rather than storing complete files, it caches only
the parts of each file that the analysis reads. For repeated passes, it can
reorganize the cached data into a compact local copy, called the ``replica
tier,'' reorganized by branch, matching the order in which an analysis reads,
and recompressed for faster access. This organization reduces the number of storage operations and
the amount of decompression required in later passes.

On suitable cache storage, the first pass, which fills the cache, adds almost
no overhead compared with reading the data directly from remote storage. If the cache encounters an
internal error, \ucache automatically falls back to forwarding requests to the
remote server, allowing the analysis to continue without depending on the
cache. By storing only the data that the analysis actually reads, rather than
complete files, a 1\,TB local cache can support a typical analysis
over datasets totaling 10--20\,TB or more.

Throughout this paper, measured quantities refer to one reference setup:
the Analysis Grand Challenge (AGC) $t\bar{t}$ analysis reading two public CMS
Open Data NanoAOD datasets, one zlib-compressed (1.8\,TB) and one
LZMA-compressed (2.4\,TB), with roughly 6\% of the bytes fetched per pass,
on machines ranging from a 6-core desktop on a residential fiber connection
to a 192-core analysis-facility node. Section~\ref{sec:method} defines the
datasets, the machines and the measurement protocol.

The remainder of this paper is organized as follows.  Section~\ref{sec:design}
describes the design of \ucache.  Section~\ref{sec:method} introduces the
benchmark analysis, the data samples, the test machines and the measurement
protocol.  Section~\ref{sec:results} presents the results, beginning with
end-to-end runtimes and then examining the mechanisms behind them. It also
compares \ucache with \xcache, the established site-level caching solution,
evaluates \ucache with ROOT's RNTuple format, the successor to TTree, and
measures how sensitive the results are to the ROOT version
(Sections~\ref{sec:xcache}, \ref{sec:rntuple}, and~\ref{sec:rootscan}).  Section~\ref{sec:limits}
discusses the limits of the approach, Section~\ref{sec:related} reviews related
work, and Section~\ref{sec:conclusions} states the conclusions. Software and data
availability, including the complete list of datasets used, is described in
Section~\ref{sec:availability} and the appendices.

\section{Design}
\label{sec:design}

The \ucache software is an XRootD \emph{client} plugin
(\texttt{libXrdClUCache.so}). To use
it, a user only needs to add a configuration file. No changes to the analysis
code are required, and nothing needs to be installed on the server. Applications
that use the XRootD client library, including ROOT, RDataFrame, uproot,
\texttt{xrdcp}, and experiment frameworks, automatically access data through the
cache. No elevated privileges are required. The plugin supports XRootD~5 clients
from version 5.6; preliminary XRootD~6 support exists in development
(Section~\ref{sec:rootscan}).

\subsection{Byte tier}
\label{sec:design-byte}

The byte tier is a local cache indexed by URL. It stores data in 4\,KiB
pages and only downloads the parts of a file that are actually requested.
Each requested range is rounded to page boundaries, fetched from the
origin, saved locally, and reused for later reads that overlap the same
range.

The cache never downloads a whole file unless the entire file is read.
In the datasets used here, a complete analysis pass fetches only
5.7--5.9\% of the dataset on the reference setup described in
Section~\ref{sec:method}.

The 4\,KiB page size determines how the cache stores and tracks data, not the
size of requests sent to the origin. Each origin request covers the full range
requested by the application, expanded to the enclosing page boundaries. This
adds at most one page at either end. On the LZMA dataset, the mean request size
is 21.8\,KiB, so rounding to 4\,KiB pages has little effect on request
size. Instead, the page size determines the granularity at which the cache
records which byte ranges are stored locally.

\begin{figure}[t]
\centering
\resizebox{\textwidth}{!}{
\begin{tikzpicture}[x=1cm,y=1cm,
  every node/.style={font=\scriptsize},
  hole/.style={pattern=north east lines, pattern color=black!30},
  unread/.style={fill=black!12},
  hdr/.style={fill=black!45},
  bJ/.style={fill=obOrange},
  bM/.style={fill=obBlue},
  bE/.style={fill=obGreen},
  frame/.style={draw=black!70, line width=0.4pt}]

  \fill[bJ] (2.2,6.05) rectangle +(0.28,0.22);
  \fill[bM] (2.56,6.05) rectangle +(0.28,0.22);
  \fill[bE] (2.92,6.05) rectangle +(0.28,0.22);
  \node[anchor=west] at (3.25,6.16) {branches read by the analysis};
  \fill[unread] (8.90,6.05) rectangle +(0.28,0.22);
  \node[anchor=west] at (9.23,6.16) {branches not read};
  \draw[frame,hole] (12.90,6.05) rectangle +(0.28,0.22);
  \node[anchor=west] at (13.23,6.16) {holes (never fetched)};

  \fill[unread] (2.5,4.4) rectangle (14.9,5.2);
  \fill[hdr] (2.2,4.4) rectangle (2.5,5.2);
  \fill[hdr] (14.9,4.4) rectangle (15.4,5.2);
  \node[anchor=south] at (2.32,5.25) {header};
  \node[anchor=south] at (15.15,5.25) {tree meta};
  \foreach \xb in {5.6,8.7,11.8} \draw[dashed,black!50] (\xb,4.4) -- (\xb,5.2);
  \foreach \i/\xc in {1/4.05,2/7.15,3/10.25,4/13.35} \node[anchor=south] at (\xc,5.25) {cluster \i};
  \foreach \xz in {2.5,5.6,8.7,11.8} {
    \fill[bJ] (\xz+0.45,4.4) rectangle +(0.14,0.8);
    \fill[bM] (\xz+1.35,4.4) rectangle +(0.11,0.8);
    \fill[bE] (\xz+2.30,4.4) rectangle +(0.09,0.8);
  }
  \draw[frame] (2.2,4.4) rectangle (15.4,5.2);
  \node[anchor=east,align=right] at (2.05,4.8) {origin file\\(cluster-major)};

  \foreach \xa in {3.02,3.905,4.845} \draw[-{Stealth},black!60] (\xa,4.35) -- (\xa,2.65);
  \node[anchor=west,align=left,text width=8.3cm] at (6.3,3.5)
    {cold pass fetches \emph{only} the requested ranges (5.9\% of this
     dataset's bytes); warm passes are served locally without contacting
     the origin};

  \draw[frame,hole] (2.2,1.8) rectangle (15.4,2.6);
  \fill[hdr] (2.2,1.8) rectangle (2.5,2.6);
  \fill[hdr] (14.9,1.8) rectangle (15.4,2.6);
  \foreach \xz in {2.5,5.6,8.7,11.8} {
    \draw[frame,bJ] (\xz+0.40,1.8) rectangle +(0.24,0.8);
    \draw[frame,bM] (\xz+1.30,1.8) rectangle +(0.21,0.8);
    \draw[frame,bE] (\xz+2.25,1.8) rectangle +(0.19,0.8);
  }
  \node[anchor=east,align=right] at (2.05,2.2) {byte cache\\\texttt{.data} (sparse)};
  \node[anchor=north east] at (15.4,1.55) {\texttt{.meta} sidecar: page bitmap $+$ CRC32C per page};

  \draw[dashed,black!50] (7.01,1.8) -- (4.0,1.05);
  \draw[dashed,black!50] (7.11,1.8) -- (11.0,1.05);
  \fill[obBlue!35] (5.0,0.35) rectangle (10.0,1.05);
  \draw[frame] (4.0,0.35) rectangle (11.0,1.05);
  \foreach \xg in {4.5,5.0,...,10.5} \draw[black!45] (\xg,0.35) -- (\xg,1.05);
  \draw[decorate,decoration={brace,amplitude=3pt}] (5.15,1.12) -- (9.80,1.12)
    node[midway,above=2pt] {logical read request};
  \node[anchor=north] at (7.5,0.28)
    {rounded to 4\,KiB pages $\cdot$ CRC32C verified per page $\cdot$ one \texttt{pread} per contiguous resident run};
\end{tikzpicture}}
\caption{Byte-tier cache layout for NanoAOD. A typical data analysis reads only a small
subset of the branches in each event cluster, producing small requests scattered
throughout the file. Requested ranges are stored in a sparse file at their
original offsets in 4\,KiB pages with per-page CRC32C checksums. Adjacent cached
pages are read together.}
\label{fig:byte-layout}
\end{figure}
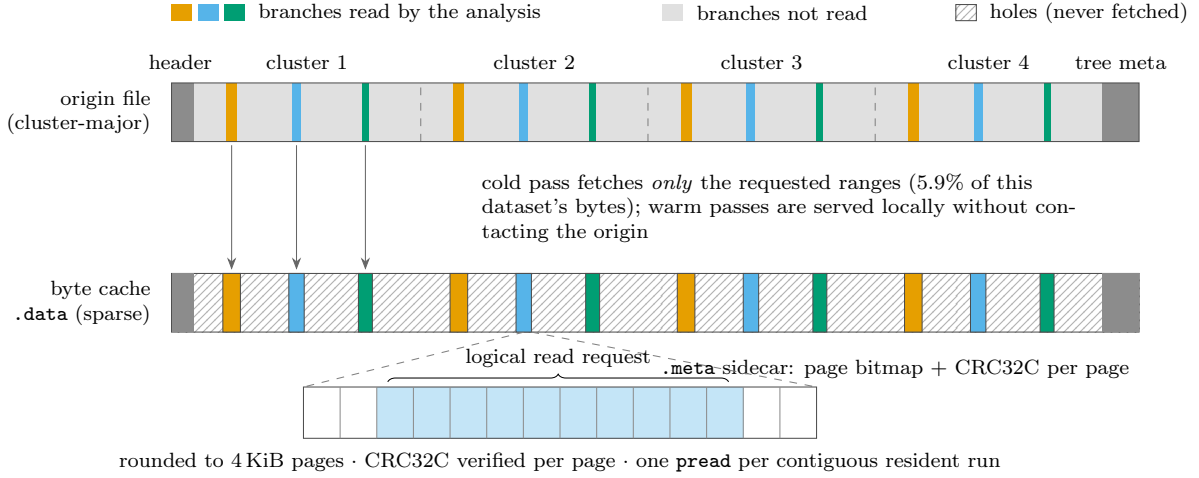

Figure~\ref{fig:byte-layout} shows the layout. Its main properties are
described below.

\begin{description}

\item[Safe fallback.] If the cache encounters an internal error, such as
a checksum mismatch, partially written metadata, or a full disk, it
fetches the requested data directly from the origin instead. Cache
failures therefore do not prevent access to the data. This fallback path
is the most heavily tested part of the code base.

\item[Data integrity.] Each cached page has a CRC32C checksum, which is
checked every time the page is read. Metadata updates are atomic, meaning
that an update is either completed fully or not applied at all. If a
process is stopped at any point, the cache will either return correct
data or fetch the affected pages again.

\item[Writing data.] Newly fetched pages are first held in memory. They
are then sorted by file offset, combined into larger blocks, and written
in operations of at most 8\,MiB. As a result, the first pass over a file
does not need to read from the cache disk and performs only a small
number of large writes. When concurrent requests need the same or
overlapping ranges, the data are fetched only once.

\item[Reading data.] When all pages needed by a request are already in
the cache, each continuous group of adjacent pages is read with one
\texttt{pread} operation. The checksum of every page is still verified
using the returned buffer. This allows the cache to read several pages
at once while continuing to check each page separately.
Section~\ref{sec:geometry} measures this behavior, showing
1.002--1.004 storage-device operations per logical request.

\item[Freshness and retries.] By default, cached data are treated as valid
for 7 days. During this period, opening a cached file does not require
contacting the origin. This default is suitable for physics datasets that
are normally written once and not modified. A separate revalidation mode
is available for data that may change. The cache can also retry failed
connections a limited number of times, with increasing delays between
attempts, to handle temporary network or origin failures.

\item[Managing storage space.] The cache starts removing data when free
disk space falls below a configured limit. It selects files using a
least-recently-used policy. The command-line interface supports cache
inspection through \texttt{ucache status}, \texttt{ucache ls}, and
\texttt{ucache stats}. It also supports removing individual or multiple
entries, preventing selected entries from being removed, checking the
configuration with \texttt{ucache doctor}, and running an end-to-end test
with \texttt{ucache test}. The \texttt{ucache bench} and
\texttt{ucache-netbench} commands measure local storage and origin
performance and are used throughout this paper.

\end{description}

\subsection{Replica tier: transpose and recompress}
\label{sec:design-replica}

\begin{figure}[t]
\centering
\resizebox{\textwidth}{!}{
\begin{tikzpicture}[x=1cm,y=1cm,
  every node/.style={font=\scriptsize},
  hole/.style={pattern=north east lines, pattern color=black!30},
  bJ/.style={fill=obOrange},
  bM/.style={fill=obBlue},
  bE/.style={fill=obGreen},
  frame/.style={draw=black!70, line width=0.4pt},
  blk/.style={draw=black!70, line width=0.4pt, font=\tiny}]

  \draw[frame,hole] (2.2,4.4) rectangle (15.4,5.2);
  \foreach \i/\xz in {1/2.5,2/5.75,3/9.0,4/12.25} {
    \node[blk,bJ,anchor=south west,minimum width=1.05cm,minimum height=0.8cm,inner sep=0] at (\xz,4.4) {J\i};
    \node[blk,bM,anchor=south west,minimum width=0.92cm,minimum height=0.8cm,inner sep=0] at (\xz+1.14,4.4) {M\i};
    \node[blk,bE,anchor=south west,minimum width=0.80cm,minimum height=0.8cm,inner sep=0] at (\xz+2.15,4.4) {E\i};
    \node[anchor=south] at (\xz+1.475,5.25) {cluster \i};
  }
  \node[anchor=east,align=right] at (2.05,4.8) {byte tier\\(origin order)};

  \draw[-{Stealth},obDark] (3.02,5.30) to[bend left=18] (6.27,5.30);
  \draw[-{Stealth},obDark] (6.27,5.30) to[bend left=18] (9.52,5.30);
  \draw[-{Stealth},obDark] (9.52,5.30) to[bend left=18] (12.77,5.30);
  \node[anchor=south,align=center] at (8.8,5.85)
    {warm read of branch J on the byte tier: one seek per cluster (mean request 21.8\,KiB on the LZMA dataset)};

  \draw[-{Stealth},line width=1.6pt,black!70] (5.0,4.25) -- (5.0,3.55);
  \node[anchor=west,align=left] at (5.25,3.9)
    {transpose $+$ recompress (LZMA-9 $\rightarrow$ ZSTD-1),\\
     built locally from already-cached bytes (background at close, or one sweep)};

  \foreach \i/\dx in {1/0.00,2/1.18,3/2.36,4/3.54}
    \node[blk,bJ,anchor=south west,minimum width=1.18cm,minimum height=0.8cm,inner sep=0] at (2.5+\dx,2.6) {J\i};
  \foreach \i/\dx in {1/0.00,2/1.03,3/2.06,4/3.09}
    \node[blk,bM,anchor=south west,minimum width=1.03cm,minimum height=0.8cm,inner sep=0] at (7.34+\dx,2.6) {M\i};
  \foreach \i/\dx in {1/0.00,2/0.90,3/1.80,4/2.70}
    \node[blk,bE,anchor=south west,minimum width=0.90cm,minimum height=0.8cm,inner sep=0] at (11.58+\dx,2.6) {E\i};
  \draw[frame] (2.5,2.6) rectangle (15.18,3.4);
  \node[anchor=east,align=right] at (2.05,3.0) {replica \texttt{.tdata}\\(branch-major)};

  \draw[decorate,decoration={brace,mirror,amplitude=3pt}] (2.5,2.5) -- (7.22,2.5)
    node[midway,below=3pt] {one sequential run per branch ($\sim$0.5\,MiB per operation)};

  \node[anchor=north west,align=left,text width=6.4cm] at (2.2,1.55)
    {\texttt{.tmeta} maps (branch, basket) $\rightarrow$ offset; reads are
     \emph{stitched}: replica where covered, byte tier otherwise; ROOT
     receives byte-identical content.};
  \node[anchor=north west,align=left,text width=6.0cm] at (9.2,1.55)
    {byte-tier pages superseded by the replica are hole-punched;
     reclamation can extend to dropping the whole original copy.};
\end{tikzpicture}}
\caption{The replica tier (schematic, not to scale). The baskets already
in the byte cache sit in the origin's cluster-major order, so a warm read
of one branch seeks once per cluster; the replica lays the same baskets
out branch-major and (optionally) transcoded LZMA-9$\rightarrow$ZSTD-1.
Measured on the LZMA dataset: the replica serves the same pass in
roughly 18 times fewer, 25 times larger operations
(Table~\ref{tab:mechanism}) with 69\% fewer instructions, at the cost of a
larger footprint: with replicas, the cache occupies 1.52 times its byte-only
size (Section~\ref{sec:opscost}).}
\label{fig:replica-layout}
\end{figure}
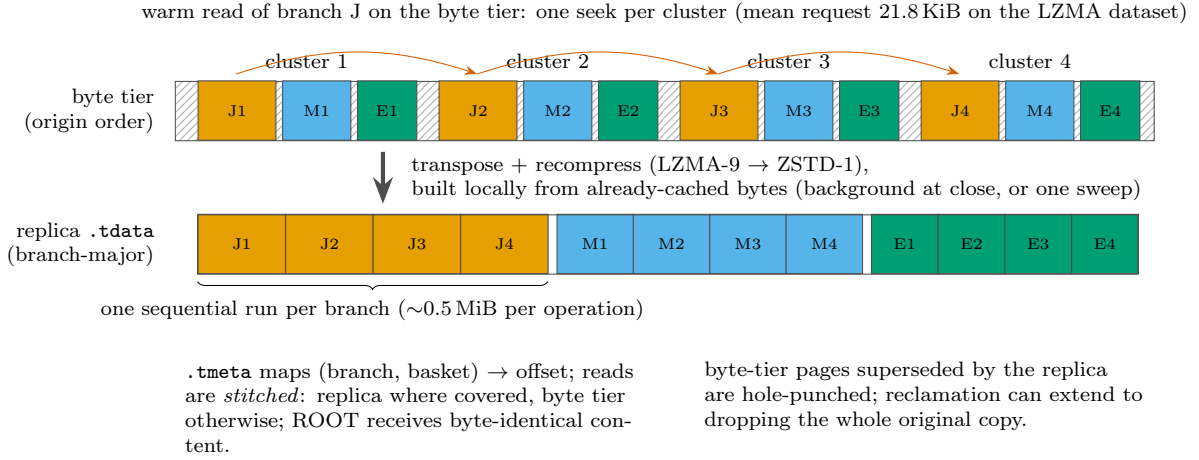

For cached ROOT files, \ucache can build a second representation entirely from
bytes already stored in the local cache, without contacting the origin. This
representation contains exactly the baskets read by the analysis and lays them
out \emph{branch-major}, with all baskets from a branch stored contiguously in
read order rather than in ROOT's event-cluster order. The baskets can also be
transcoded from the source codec to ZSTD level~1. Serving remains transparent:
ROOT reads the same URLs and receives byte-identical logical content, with each
request assembled from the replica where available and from the byte tier
otherwise. Figure~\ref{fig:replica-layout} illustrates replica tier layout. This
representation serves two purposes:

\begin{itemize}
\item \textbf{Geometry}: in the origin layout, the baskets for a given branch
  are spread across the file, with one basket in each event cluster. The data
  between them mostly belongs to branches that the analysis does not read. As a
  result, reading one branch requires many small, separate reads. Branch-major
  storage instead places the baskets for each branch next to each other, so a
  single larger read can retrieve several baskets at once. On the LZMA dataset,
  the replica tier serves the same pass with roughly 18 times fewer and
  25 times larger reads (Section~\ref{sec:geometry}).
\item \textbf{Codec}: strong compression reduces the storage footprint of the
  original dataset, but codecs such as LZMA, and even zlib, can be expensive to
  decompress and may become a bottleneck in a fast analysis loop. The replica
  can therefore transcode cached baskets to ZSTD-1, trading a modest increase in
  size for much faster decompression. For LZMA-9 sources, the default
  compression algorithm for CMS NanoAOD, moving decoding to ZSTD-1 removes most
  of the decompression cost (the byte-tier pass executes 3.2 times as many
  instructions as the replica pass, Section~\ref{sec:codec}) at the cost of a
  larger on-disk footprint (Section~\ref{sec:opscost}). Transcoding is done once, when the replica
  is built, while subsequent analysis passes benefit from the faster codec.
\end{itemize}

Replica builds can run either as a foreground sweep (\texttt{ucache recompress})
or automatically in the background. With \texttt{recompress~=~on}, files are
queued when they are closed and processed opportunistically by a low-priority
background process, so recompression uses spare CPU resources without
interfering with normal analysis work. The amount of parallel work is adjusted
to the host.

Three constraints prevent background replica builds from interfering with normal
cache operation.

\begin{description}

\item[Free space.] A replica is created before the byte-tier data it replaces is
  released, so a build temporarily increases cache usage. Each file is therefore
  checked against the same free-space threshold used for eviction. After a build
  completes, the superseded byte-tier pages are released, making space available
  for the next file. If there is not enough space, the file is deferred and
  re-queued until space becomes available. A foreground sweep behaves
  differently: it estimates the total temporary growth before starting and asks
  the user for confirmation if needed.

\item[Time.] Only one build pass can run at a time, because concurrent passes
  could reclaim pages that another build is still using. Background builds
  therefore run only until a deadline, then re-queue any remaining files. This
  limits how long a user-initiated foreground sweep must wait. The plugin
  periodically restarts the background pass, allowing a partially processed
  queue to continue making progress.

\item[Verification.] Replica builds read their input from the byte tier, where
  some pages may be absent because the analysis never requested them or because
  they have already been reclaimed. Missing pages read back as zeros, so every
  page is checked against its stored CRC32C checksum before use. If a file
  contains missing pages, the build is re-queued as \emph{incomplete} rather
  than marked as failed, and can succeed once those pages are
  available. Genuinely malformed input is treated as a hard failure.

\end{description}

Once a replica is published and validated, the byte-tier pages it supersedes
are punched out of the sparse cache file. Configuring reclaim as
\texttt{full} instead releases the file's entire byte-tier copy, which gives
the smallest steady-state footprint; any read the replica does not cover is
then refetched from the origin. Section~\ref{sec:opscost} quantifies these
operational costs.

\section{Benchmarking method}
\label{sec:method}

While running, \ucache records detailed information about its activity,
including the number of bytes and read operations served by each cache tier,
the distributions of read sizes and latencies, and a record for every file it
handles. The measurements in this paper use these counters in addition to
wall-clock time. When an analysis pass becomes faster or slower, the counters
show where the data came from and how the storage system behaved, making the
cause of the change easier to identify.

\subsection{Workload and datasets}

\begin{table}[t]
\centering\footnotesize
\setlength{\tabcolsep}{4pt}
\caption{The two benchmark datasets. ``Fetched'' is the volume
\ucache fetches to serve the analysis, after rounding requests to whole
cached pages. Compression ratios are measured
from the files.}
\label{tab:datasets}
\begin{tabular}{lllrrrr}
\toprule
Dataset & Source & Algorithm (ratio) & Files & Volume & \makecell{Events\\($10^9$)} & Fetched \\
\midrule
\datasetZ & AGC NanoAOD (2015 Open Data) & zlib-1 (2.9--3.3) & 787 & 1.78\,TB & 0.94 & 101\,GB (5.7\%) \\
\datasetL & Open Data UL2016 NanoAODv9 & LZMA-9 (3.7--6.0) & 1456 & 2.41\,TB & 1.30 & 143\,GB (5.9\%) \\
\bottomrule
\end{tabular}
\end{table}

The workload is the AGC CMS Open Data
$t\bar{t}$ analysis~\cite{agc-repo,shadura2023agc}, using its RDataFrame
implementation. It is a realistic analysis that performs object selections,
systematic variations, and histogramming while reading 15 NanoAOD branches.
Its 9 sample/variation pairs are processed as 9 concurrent RDataFrame graphs.
ROOT's \texttt{TTreeCache} is enabled, so \ucache sits below a client that
already batches and prefetches reads. The gains reported here therefore come
on top of ROOT's own prefetching rather than replacing it.

We use 2 datasets to cover the 2 compression regimes relevant to this
study (Table~\ref{tab:datasets}). The first is the AGC team's 2015-derived
NanoAOD dataset, compressed with zlib-1, which is relatively inexpensive to
decompress. The second maps the same analysis onto official CMS Open Data
UL2016 NanoAODv9 records compressed with LZMA-9, the production NanoAOD
codec, which is considerably more expensive to decompress. We selected 26
published records to mirror the AGC samples role by role, without changing
the analysis code (Appendix~\ref{app:datasets}).

In both datasets, the analysis reads only about 6\% of the stored data. A
complete pass fetches 101\,GB from the 1.8\,TB zlib dataset and 143\,GB from
the 2.4\,TB LZMA dataset.

\subsection{Test machines}

We run the benchmark on 6 machines chosen to represent the environments
\ucache is intended to support: a personal desktop, a workstation located close
to the data source, 2 analysis-facility machines (one adjacent to a major
computing center and one across the wide-area network), a cloud virtual
machine, and an Apple-silicon system representative of a typical laptop.
Table~\ref{tab:fleet} lists their hardware.

\begin{table}[t]
\centering\footnotesize
\setlength{\tabcolsep}{4pt}
\caption{Test machines and cache configurations. SSD models named without a
maker are Samsung. Machines with 2 storage rows were tested with the cache
on each configuration. Source round-trip times are measured during each run, together
with jitter and packet loss. \mD\ and \mM\ share the same location and fiber
connection, so differences between them reflect the machines rather than the
network.}
\label{tab:fleet}
\resizebox{\textwidth}{!}{\begin{tabular}{lllrll}
\toprule
Tag & Machine & CPU & \makecell[c]{RAM\\GB} & Cache storage & Source link \\
\midrule
\mD & desktop & Ryzen 9600X, 6 cores & 16 & \makecell[tl]{1\,TB SATA SSD (860 EVO)\\1\,TB NVMe SSD (990 PRO)} & fiber, 17.5\,ms \\
\mW & workstation & 2$\times$Xeon 4216, 32 cores & 192 & 2\,TB SATA SSD (PM883) & CERN LAN, 0.3\,ms \\
\mFone & facility (MIT) & 2$\times$EPYC 9654, 192 cores & 1500 & 2$\times$1\,TB NVMe SSD (PM9A3) & CERN 78, FNAL 26\,ms \\
\mFtwo & facility (CERN) & EPYC 9655P, 96 cores & 768 & 7.7\,TB NVMe SSD (Micron 7450) & CERN LAN, 0.2\,ms \\
\mV & cloud VM & EPYC Genoa, 24 cores & 91 & \makecell[tl]{VM-local NVMe (virtualized)\\network volume (rate-limited)} & CERN LAN, 0.3\,ms \\
\mM & Mac mini & Apple M2 Pro, 10 cores & 32 & 1\,TB internal NVMe SSD & fiber, 18.1\,ms \\
\bottomrule
\end{tabular}}
\end{table}

Table~\ref{tab:bench} characterizes each cache device using the storage
self-test included with \ucache (\texttt{ucache bench}). Except where
noted for macOS, the test uses \texttt{O\_DIRECT}, a 69\,GB test file, and
60\,s measurements. The self-test reports two groups of measurements.

The first group, \emph{Standard}, uses the same access patterns on every
machine and can be compared directly with device specifications. It measures
sequential read bandwidth, in-place write bandwidth, and random 4\,KiB reads
at queue depths (QD) 1 and 16, reported as input/output operations per
second (IOPS). The change between the two queue depths helps
distinguish a storage device that benefits from parallel requests from one
limited by an operations-per-second quota. A healthy device improves by
9--16 times, while a rate-limited device remains nearly flat.

The second group measures \ucache's \emph{own code} at the same concurrency
used by the analysis. It first measures a cold fill through the normal
\ucache write path, including staging, offset-sorted drains, and checksum
generation. A cold analysis pass is primarily a write workload for the cache,
because pages staged in memory can be returned to the client before they are
written to storage. The test then drops the operating system page cache and
reads the complete test file back through \ucache using two access patterns:
scattered $\sim$48\,KiB reads, similar to the byte tier, and sequential
$\sim$512\,KiB reads, similar to the replica tier. The benchmark models only
the order in which offsets arrive.

These measurements provide the storage limits used to interpret the analysis
results in Section~\ref{sec:codec}. The benchmark that runs through \ucache's
actual write path allows us to assess cache suitability without benchmarking a
complete analysis. Every storage test was run with the device otherwise idle,
and block-layer counters show that more than 98\% of the device traffic during
each run came from the self-test.

The six machines are described below.

\paragraph{\mD: desktop.}
This machine represents a typical use case for an individual physicist working on
personal hardware over a home network or on an office computer.

\begin{itemize}
\item CPU: AMD Ryzen 5 9600X (Zen 5, 2024), with 6 cores and 12 simultaneous
multithreading (SMT) threads. It has relatively few but fast cores. The benchmark runs with 6 and
12 threads.

\item RAM: 16\,GB, smaller than every cached dataset. The operating
system's page cache therefore cannot hold the complete working set.

\item Cache device: 1\,TB SATA SSD (Samsung 860 EVO), a consumer drive whose
single-level-cell (SLC) write buffer holds up to 42\,GB. A cold fill writes far more than that,
but unlike the NVMe drive below this device shows no burst-versus-sustained
cliff: its rated write rate falls only from 520 to 500\,MB/s once the buffer
is exhausted, and a 69\,GB test file written over 130\,s measured 529\,MB/s
in both the first and the last quarter of that window. The measured in-place
rate of 530\,MB/s is therefore a sustained rate. All measured read patterns
reach 0.48--0.57\,GB/s, close to the SATA interface limit, so the byte and
replica tiers have similar read bandwidth on this device.

\item Second cache device: 1\,TB NVMe SSD (Samsung 990 PRO, PCIe~4.0), a
consumer drive with a $\sim$115\,GB SLC write buffer and a rated read
bandwidth of 6.9\,GB/s. Once the write buffer is exhausted, writes fall to
the sustained rate of the flash itself. The measured in-place write rate is
2.5\,GB/s. Because the benchmark cycles over a 69\,GB file for 60\,s, the
measurement extends beyond the temporary benefit of the write buffer.
Sequential reads reach 6.3\,GB/s, while random 4\,KiB reads reach 291k
operations per second at queue depth 16.

\item Network: residential fiber, with 17.5\,ms latency to the CERN source.
\end{itemize}

\paragraph{\mW: CERN workstation.}
This machine represents a server-class system with the data source already
available on the local network. It is similar to machines in the \texttt{lxplus}
analysis cluster. In general, this is a difficult case for caching because of
its close proximity to the source.

\begin{itemize}
\item CPU: 2$\times$ Intel Xeon Silver 4216 (Cascade Lake, 2019), for a
total of 32 cores at 2.1\,GHz base frequency. The machine has many cores,
but each core is relatively slow. The benchmark runs with 32 threads.

\item RAM: 192\,GB.

\item Cache device: 1.92\,TB SATA SSD (Samsung PM883), a data center drive
without an SLC write buffer. Its rated performance is therefore sustained by
design: 520\,MB/s for writes and 550\,MB/s for reads. The measured in-place
write rate is within 3\% of the specification across runs. As on the
desktop's SATA SSD, all read patterns approach the SATA interface limit, so
the byte and replica tiers provide similar read bandwidth.

\item Network: CERN LAN, with 0.30\,ms latency to the source.
\end{itemize}

\paragraph{\mFone: subMIT analysis-facility
node~\cite{dalfonso2026submit}.} This machine represents the large-scale
analysis case, with storage much faster than the workload requires.

\begin{itemize}
\item CPU: 2$\times$ AMD EPYC 9654 (Zen 4, 2022), for a total of 192
cores. The benchmark runs with 64 threads.

\item RAM: 1.5\,TB.

\item Cache device: two 960\,GB NVMe SSDs (Samsung PM9A3, data center
class) behind a logical volume on the system disk. Each drive is rated at
6.5\,GB/s for reads, 1.5\,GB/s for sustained writes, and 580k random
reads/s. Measured read performance is close to the specification of a
single drive: 5.5\,GB/s for sequential reads and up to 328k random reads/s
at queue depth 32. This is roughly 10 times the read performance of the
two SATA systems.

The write path is considerably slower. In-place writes reach 564\,MB/s at
queue depth 1, while a \ucache fill with 4 writers reaches 895\,MB/s.
Both are well below the rated write speed of a single drive. The host also
limits the amount of dirty buffered data to 100\,MB
(\texttt{vm.dirty\_bytes}), so writes generated during a fill reach the
storage device with little buffering.

\item Network: wide-area, with 78\,ms latency to CERN and 26\,ms to
Fermilab (FNAL).
\end{itemize}

\paragraph{\mFtwo: CERN batch node.} This machine represents
current-generation facility hardware adjacent to the data source. The
measurements ran as batch jobs submitted through the CMS global pool to CERN
worker nodes, inside the standard container environment with a grid proxy,
the setting in which an physicist would actually use such hardware. Jobs land
where the pool schedules them: the analysis campaigns landed on 2 nodes of
the same model, whose raw storage patterns agree within 1\%, and the storage
characterization below ran on a third; we treat the class as one machine.

\begin{itemize}
\item CPU: AMD EPYC 9655P (Zen 5, 2024), with 96 cores and 192 SMT threads.
The benchmark runs with 64 threads inside a batch slot; the remaining cores
serve other jobs, whose load is not under our control. Unlike the other 5
machines, this one is therefore never measured idle.

\item RAM: 768\,GB.

\item Cache device: a partition covering nearly the whole of the node's 7.7\,TB NVMe system
disk (Micron 7450, data center class). Sequential reads reach 4.2\,GB/s at
queue depth 1 and 6.1\,GB/s at queue depth 64, and random 4\,KiB reads scale
from 14k operations per second at queue depth 1 to 418k at queue depth 32.
Through \ucache's own paths at the analysis concurrency, a fill writes at
3.0\,GB/s and reads reach 6.5\,GB/s in the replica tier's sequential shape
(matching the raw device) but 2.6\,GB/s in the byte tier's scattered
48\,KiB shape, 47\% of the 5.4\,GB/s the raw device delivers for that same
pattern: the byte path achieves a queue depth of only 12 while 64 threads
are asking.

\item Network: CERN LAN, with 0.2\,ms latency to both EOS sources used,
through a 10\,GbE interface with standard 1500-byte frames (measured on
3 nodes of the class).
\end{itemize}

\paragraph{\mV: cloud virtual machine.} This machine represents a cloud
worker node of the latest generation. We test 2 storage configurations on the
same VM: a fast VM-local NVMe device and a slower, rate-limited network block
volume.

\begin{itemize}
\item CPU: AMD EPYC (Genoa, virtualized), with 24 virtual cores. The
benchmark runs with 24 threads; the \xcache comparison on this machine
(Section~\ref{sec:xcache}) used 16.

\item RAM: 91\,GB, also smaller than the cached datasets.

\item Cache device, first configuration: VM-local NVMe. Because the device is
  virtualized, the physical drive model is not visible from inside the
  guest. Its measured bandwidth requires an important qualification:
  \texttt{O\_DIRECT} bypasses the guest's page cache, but not the hypervisor's
  cache. Replica-shaped reads reach 9.8\,GB/s through \ucache's own path and up
  to 12.7\,GB/s in the raw test, higher than this hardware class can deliver
  from storage alone. The virtual disk was therefore serving some data from host
  memory. These bandwidth measurements should be treated as upper bounds on the
  virtual device. The measured latency, in the tens of microseconds, and
  approximately 100k random reads per second are more representative of the
  storage itself.

\item Cache device, second configuration: a Ceph-backed network block volume,
  the standard cloud storage option provided with a fixed operations-per-second
  limit. The underlying Ceph cluster is large and shared, but the service class
  applies limits at the virtual disk presented to the VM. Reads and writes are
  each limited at about 1000 operations per second, with a bandwidth ceiling of
  a few hundred MB/s.

The self-test makes this limit clear. Random-read performance remains at about
1000 operations per second at every queue depth, whereas a local storage device
normally improves by roughly 9--16 times as concurrency increases. Median
latency rises from 0.9 to 32\,ms as queue depth increases.  Additional requests
therefore wait in a queue rather than being served in parallel.

Sequential bandwidth also varies with the load on the shared service. The
two most recent measurements reached 153 and 158\,MB/s, compared with
315\,MB/s in the first measurement campaign. A small, fast local disk
alongside a larger, rate-limited network volume is a common configuration
for worker nodes.

\item Network: CERN LAN, with 0.3\,ms latency to the source.
\end{itemize}

\paragraph{\mM: Apple Mac mini.} This machine represents a typical laptop
or small office desktop, similar to \mD, but with a different processor
architecture and operating system. It is connected to the same network
at the same location, so differences between \mM\ and \mD\ reflect the
machines rather than the network connection.

\begin{itemize}
\item CPU: Apple M2 Pro (2023), with 10 heterogeneous cores: 6 performance
cores and 4 efficiency cores. The benchmark runs with 6 and 10 threads.
macOS does not provide a mechanism to pin threads to one core class, so the
reported thread count specifies how many threads run, not which cores run
them.

\item RAM: 32\,GB, smaller than every cached dataset.

\item Cache device: the machine's internal 1\,TB NVMe SSD (Apple AP1024Z).
It is also the system disk, which reflects the storage an physicist would
typically use on a personal desktop. Its measurements are reported in
Table~\ref{tab:bench}. The test differs from those on the other machines in
one respect: macOS does not provide \texttt{O\_DIRECT}, so the benchmark
uses \texttt{F\_NOCACHE} instead to keep reads out of the operating system
page cache. Unlike \texttt{O\_DIRECT}, \texttt{F\_NOCACHE} is a request
rather than a strict guarantee. The measured latency supports that it is
working as intended: 0.06\,ms at queue depth 1 is consistent with
storage-device latency rather than a memory-cache hit.

\item Network: the same residential fiber connection as \mD, with
18.1\,ms latency to the CERN source.
\end{itemize}

\begin{table}[t]
\centering\small
\setlength{\tabcolsep}{4pt}
\caption{Measured cache-device performance. Standard access patterns: sequential
  4\,MiB reads and in-place writes at queue depth 1, random 4\,KiB reads at
  queue depths 1 and 16, and median latency at queue depth 1. The last 3
  columns exercise uCache's own fill and read paths at the thread count $N$ used
  for analysis on each machine.  These measurements therefore reflect both the
  device and uCache's own I/O paths, at the release current when each machine
  was characterized (Appendix~\ref{app:repro}). $^{\ast}$The guest cannot bypass the
  hypervisor's cache.  $^{\dagger}$\mM\ runs macOS, where the tool requests
  rather than guarantees page-cache bypass.}
\label{tab:bench}
\begin{tabular}{lrrrrrrrrr}
\toprule
 & & \multicolumn{4}{c}{Standard (fixed shapes)} & & \multicolumn{3}{c}{Through uCache at $N$ threads} \\
\cmidrule(lr){3-6}\cmidrule(lr){8-10}
Cache device & $N$ & \makecell{seq read\\MB/s} & \makecell{seq write\\MB/s} & \makecell{4\,KiB read\\QD1$\to$16, IOPS} & \makecell{QD1 p50\\ms} & & \makecell{cold fill\\MB/s} & \makecell{byte read\\MB/s} & \makecell{replica read\\MB/s} \\
\midrule
\mD\ SATA  & 6  & 559 & 530   & 11k $\to$ \z99k  & 0.09 & & 465  & 476  & 560 \\
\mD\ NVMe  & 6  & 6323 & 2492 & 18k $\to$ 291k & 0.05 & & 1499 & 1637 & 3471 \\
\mW\ SATA  & 32 & 555 & 508   & \z9k $\to$ \z94k & 0.11 & & 309  & 545  & 562 \\
\mFone\ NVMe  & 64 & 5490 & 564  & 15k $\to$ 199k & 0.06 & & 895  & 2775 & 4921 \\
\mFtwo\ NVMe & 64 & 4217 & 3938 & 14k $\to$ 220k & 0.06 & & 2964 & 2566 & 6528 \\
\mV\ NVMe$^{\ast}$ & 16 & 5956 & 3043 & \z7k $\to$ \z95k & 0.14 & & 1654 & 2225 & 9837 \\
\mV\ cloud & 16 & 153 & 154   & \z1k $\to$ \z\z1k & 0.88 & & 129  & 40   & 188 \\
\mM\ NVMe$^{\dagger}$ & 6  & 4874 & 3545 & 15k $\to$ 182k & 0.06 & & 570  & 1116 & 4048 \\
\mM\ NVMe$^{\dagger}$ & 10 & 4586 & 3288 & 16k $\to$ 183k & 0.06 & & 615  & 1848 & 3689 \\
\bottomrule
\end{tabular}
\end{table}

\subsection{Remote data sources}

Both datasets are served by the CERN Open Data EOS instance
(\texttt{eospublic.cern.ch}), a public shared service. \datasetL has a
second source: it was replicated to Fermilab and read there through that
site's dCache endpoint (\texttt{cmsdcadisk.fnal.gov}). \datasetZ was read
only from CERN. Round-trip times from each machine are listed in
Table~\ref{tab:fleet}.

Both services are shared production systems, so what either one delivers to
a single client depends on the load other users place on it at the time.
Section~\ref{sec:origins} measures how much this varies, and which
comparisons stay meaningful in spite of it.

\subsection{Protocol}
\label{sec:protocol}

Each benchmark run starts from a clean cache. Before timing begins, a
precondition step records and removes any state left by earlier runs. The
following measurements are then performed in order:

\begin{enumerate}
\item \textbf{direct}: the plugin is loaded but caching is disabled. This
provides the no-cache baseline using the same binaries and the same session;

\item \textbf{cold}: the first pass through an empty cache, which fills the
byte tier. Our design target is for this pass to take no more than
1.1 times the direct-read time, so enabling the cache adds little initial
cost. Exceeding this threshold does not invalidate a run. Instead, all cold
results are evaluated against the same 1.1-times target, including systems
whose storage cannot meet it (Section~\ref{sec:limits});

\item \textbf{warm-byte}: a second pass served from the byte tier. Before
this measurement, cached pages are evicted from the operating system page
cache, and the eviction is verified page by page with \texttt{mincore}. The
exception is \mM: macOS provides no selective page-cache eviction, so its
warm passes run without it; its working sets exceed the machine's RAM. We
also repeat the pass without eviction, reported as the \textbf{RAM} passes, to
measure the benefit of keeping cached data in memory. RAM-resident results
are not reported as storage performance;

\item \textbf{cold + recompress-on}: another fill from an empty cache, this
time with background replica building enabled. This measures how much the
background recompression work interferes with the fill;

\item \textbf{sweep}: a foreground \texttt{ucache recompress} pass that
builds replicas for all eligible cached data;

\item \textbf{warm-replica}: a repeat pass served from the replica tier,
again measured both with and without operating system page-cache eviction.
\end{enumerate}

Every timed measurement is checked against counters reported by the plugin
itself. For warm passes, we record the \emph{origin share}, the fraction of
cache-served bytes that still had to be fetched from the remote source into
the cache. An origin share of
$\le$0.1\% passes the check, while values up to $\le$5\% generate a
warning. Files the cache declines to admit, which happens only when the
working set exceeds the cache (Section~\ref{sec:limits}), are relayed
from the origin without being stored; their volume is reported separately,
because relaying is the designed behavior of an over-run cache rather than
a cache miss. We also require CRC failures $=$ 0, fail-open events $=$ 0, and
validation failures $=$ 0. For every warm-replica measurement, we report the
\emph{replica coverage} alongside the wall-clock time. This is necessary
because a partially replicated dataset mixes byte-tier and replica reads, and
our measurements show that partial coverage provides little benefit. We
therefore do not quote a warm-replica result without its coverage.

For all runs reported in this paper, CRC failures, fail-open events, and
validation failures were zero on every measurement. Warm-pass origin share was
exactly 0 on every measurement with the default reclaim setting; the
deliberately over-run campaign of Section~\ref{sec:limits} additionally
relayed a fixed 12.7\% of bytes on every warm pass, for the files it
declined to admit. The
Mac mini campaigns, which used \texttt{recompress\_reclaim = full},
refetch about 0.01\% of served bytes on warm-replica passes
(Section~\ref{sec:opscost}).

Three measurement rules proved important in obtaining reproducible results:

\begin{itemize}
\item \textbf{Origin measurements are comparable only within the same run.}
Throughput from the shared remote source can vary by 2--4 times within
an hour (Section~\ref{sec:origins}). Comparing cold or direct measurements
from different sessions can therefore measure changes in source load rather
than changes caused by the cache. Warm measurements are local and are much
more stable, reproducing across days and software versions to
about 1\%.

\item \textbf{Evicted and RAM-resident warm passes measure different things.}
On the workstation, the \datasetZ byte tier takes 230.2\,s after operating
system page-cache eviction, compared with 131.9\,s when the same data remains
resident in memory. Only the evicted and verified measurements are used to
characterize storage performance. Quoting evicted walls also bounds memory
pressure: every quoted warm result already assumes the page cache
contributes nothing, so competition for memory from the analysis itself
cannot make the cache-serving side slower than reported.
\end{itemize}

Wall-clock time is separated into ROOT's graph-construction phase
(\emph{building}), which includes opening every file, and the parallel event
loop (\emph{executing}). This distinction matters because graph construction
does not benefit from additional analysis threads and is sensitive to remote
file-open latency. Combining the two phases into a single ratio can therefore
distort comparisons, especially when one execution phase is much faster than
another (Section~\ref{sec:openlatency}).

Instruction counts and instructions per cycle (IPC) are collected using
aggregate \texttt{perf\_event\_open} counters. These counters measure
user-mode execution only; kernel time is not sampled, and none of the results
below relies on attributing work to the kernel.

\paragraph{External validation.}
The subMIT facility study~\cite{dalfonso2026submit} independently measured
the same analysis on the same class of machine, with 192 cores and 1.5\,TB
of RAM, using a different ROOT version several months earlier. From its
Figure~6, the remote XRootD measurement from CERN is 335\,s, compared with
328--374\,s for our direct and cold measurements. Its local-storage
graph-construction time is 9.0\,s, compared with 9.17\,s when our analysis is
served from the cache. These independently obtained measurements agree at the
few-percent level and provide an external check that our benchmark setup
produces results consistent with an earlier study.

\section{Benchmarking results}
\label{sec:results}

This section evaluates the overall performance of \ucache and examines it in
detail under various conditions and on different hardware. The goal is to better
understand the factors contributing to faster analysis and to provide
recommendations for optimal usage. One reading rule applies throughout:
speedup ratios divide by direct-reading baselines that move with the load on
the shared remote sources, so the absolute warm-cache times, which reproduce
to about one percent, are the stable quantity
(Section~\ref{sec:origins} quantifies both). It also compares \ucache with \xcache,
measures the cache on ROOT's RNTuple format, and repeats the TTree
measurement across 4 ROOT versions.

\subsection{Performance overview}
\label{sec:endtoend}

\begin{figure}[t]
\centering
\includegraphics[width=\textwidth]{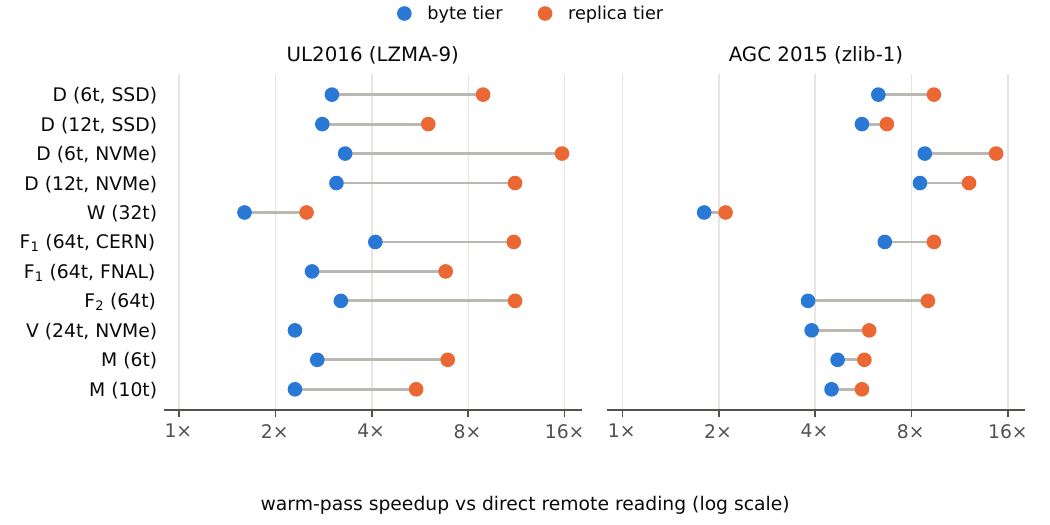}
\caption{Warm-pass speedups of Tables~\ref{tab:e2e-lzma}
  and~\ref{tab:e2e-zlib}: each warm wall divided into its own campaign's
  direct wall, per machine configuration and cache tier, on a logarithmic
  axis. A connector spans one configuration's byte-tier and replica-tier
  speedups. The replica tier is absent on \mV\ for \datasetL\ (its store
  does not fit that machine's cache disk), \datasetZ\ was read from CERN
  only, and the rate-limited cloud volume is excluded
  (Section~\ref{sec:limits}).}
\label{fig:speedups}
\end{figure}

The first practical question is how much the cache reduces analysis time, and
whether building the cache makes the first pass slower.
Tables~\ref{tab:e2e-lzma} and~\ref{tab:e2e-zlib} show the results for the two
datasets, which use different compression algorithms, and
Figure~\ref{fig:speedups} summarizes their warm-pass speedups.

\begin{table}[t]
\centering\small
%
%
\caption{End-to-end wall-clock time in seconds for \datasetL (LZMA-9). Ratios
  compare each warm measurement with the direct time in the same row.
  \emph{Fill overhead} is the measured wall-clock cost of filling the cache
  during the first pass, as a percentage of the pass (see
  text). \emph{Replica rate} is the local read rate during
  the warm-replica pass. \emph{Speedup} compares warm-byte with warm-replica and
  captures the combined benefit of recompression and data
  layout. All rows were measured with release 0.19.1, except the \mM\ rows
  (release 0.20.0) and the \mV\ row (release 0.21.0). Replicas cover all 1456 files; on \mV\ the replica
  store for this dataset does not fit on the machine's cache disk, so only
  the byte tier is reported.}
\label{tab:e2e-lzma}
\begin{tabular}{llrrr@{\,}rr@{\,}rrr}
\toprule
Machine & Source & Direct & \makecell{Fill\\overhead} & \multicolumn{2}{c}{Warm-byte} & \multicolumn{2}{c}{Warm-replica} & \makecell{Replica rate\\MB/s} & Speedup \\
\midrule
\mD\ (6t, SSD)  & CERN & 3237 & 0.1\% & 1065 & (3.0$\times$) & 362 & (8.9$\times$) & 538 & 2.94 \\
\mD\ (12t, SSD) & CERN & 1938 & 0.1\% & 692 & (2.8$\times$) & 326 & (6.0$\times$) & 598 & 2.13 \\
\mD\ (6t, NVMe) & CERN & 3362 & 0.1\% & 1008 & (3.3$\times$) & 215 & (15.7$\times$) & 906 & 4.70 \\
\mD\ (12t, NVMe) & CERN & 1974 & 0.1\% & 647 & (3.1$\times$) & 176 & (11.2$\times$) & 1108 & 3.68 \\
\mW\ (32t) & CERN & 856 & 0.4\% & 544 & (1.6$\times$) & 339 & (2.5$\times$) & 579 & 1.61 \\
\mFone\ (64t) & CERN & 621 & 0.3\% & 152 & (4.1$\times$) & 56 & (11.1$\times$) & 3537 & 2.71 \\
\mFone\ (64t) & FNAL & 390 & 3.1\% & 152 & (2.6$\times$) & 57 & (6.8$\times$) & 3466 & 2.65 \\
\mFtwo\ (64t) & CERN & 472 & 0.1\% & 149 & (3.2$\times$) & 42 & (11.2$\times$) & 4713 & 3.55 \\
\mV\ (24t, NVMe) & CERN & 997 & 0.3\% & 434 & (2.3$\times$) & -- & & -- & -- \\
\mM\ (6t) & CERN & 3835 & 1.2\% & 1446 & (2.7$\times$) & 559 & (6.9$\times$) & 346 & 2.58 \\
\mM\ (10t) & CERN & 2500 & 1.2\% & 1111 & (2.3$\times$) & 456 & (5.5$\times$) & 424 & 2.44 \\
\bottomrule
\end{tabular}
\end{table}

\begin{table}[t]
\centering\small
\caption{End-to-end wall-clock time in seconds for \datasetZ (zlib-1). Ratios
  compare each warm measurement with the direct time in the same row.
  \emph{Fill overhead} is the measured wall-clock cost of filling the cache
  during the first pass, as a percentage of the pass; for the cloud-volume
  row it is instead the measured difference between the first pass and a
  direct read, relative to the direct time (see text). \emph{Replica rate} is the local read rate
  during the warm-replica pass. \emph{Speedup} compares warm-byte with
  warm-replica and captures the combined benefit of recompression and data
  layout. All rows were measured with release 0.19.1, except the \mM\ rows
  (release 0.20.0).}
\label{tab:e2e-zlib}
\begin{tabular}{llrrr@{\,}rr@{\,}rrr}
\toprule
Machine & Source & Direct & \makecell{Fill\\overhead} & \multicolumn{2}{c}{Warm-byte} & \multicolumn{2}{c}{Warm-replica} & \makecell{Replica rate\\MB/s} & Speedup \\
\midrule
\mD\ (6t, SSD) & CERN & 1989 & 0.2\% & 316 & (6.3$\times$) & 212 & (9.4$\times$) & 487 & 1.49 \\
\mD\ (12t, SSD) & CERN & 1242 & 0.1\% & 223 & (5.6$\times$) & 186 & (6.7$\times$) & 556 & 1.20 \\
\mD\ (6t, NVMe) & CERN & 1975 & 0.2\% & 224 & (8.8$\times$) & 134 & (14.7$\times$) & 771 & 1.67 \\
\mD\ (12t, NVMe) & CERN & 1297 & 0.1\% & 152 & (8.5$\times$) & 107 & (12.1$\times$) & 964 & 1.41 \\
\mW\ (32t) & CERN & 420 & 0.5\% & 230 & (1.8$\times$) & 199 & (2.1$\times$) & 525 & 1.16 \\
\mFone\ (64t) & CERN & 349 & 0.8\% & 53 & (6.6$\times$) & 37 & (9.4$\times$) & 2864 & 1.41 \\
\mFtwo\ (64t) & CERN & 262 & 0.1\% & 69 & (3.8$\times$) & 29 & (9.0$\times$) & 3676 & 2.37 \\
\mV\ (24t, NVMe) & CERN & 506 & 0.5\% & 130 & (3.9$\times$) & 86 & (5.9$\times$) & 1212 & 1.52 \\
\mV\ (24t, cloud) & CERN & 482 & 514.6\% & 3321 & (0.1$\times$) & 1451 & (0.3$\times$) & 72 & 2.29 \\
\mM\ (6t) & CERN & 1951 & 0.9\% & 415 & (4.7$\times$) & 345 & (5.7$\times$) & 303 & 1.20 \\
\mM\ (10t) & CERN & 1430 & 0.8\% & 318 & (4.5$\times$) & 254 & (5.6$\times$) & 410 & 1.25 \\
\bottomrule
\end{tabular}
\end{table}

Replica coverage for \datasetZ was complete on every machine, including
both cloud-VM configurations. The rate-limited cloud volume has a strict
limit on the number of I/O operations it can perform; on it, both cache
tiers ran slower than direct access (the byte cache at 0.15 and the replica
cache at 0.33 times the direct speed). Section~\ref{sec:geometry} examines
this volume.

The \emph{fill overhead} column measures how much time is added while the cache
is being built during the first pass. This value comes from counters inside the
plugin rather than from comparing 2 separate runs. While data are being read,
the plugin places data waiting to be written into a staging buffer. A reading
thread has to stop only when this buffer is full and must wait for some data to
be written out. The plugin records the total time spent in these waits. Dividing
this total by the thread count estimates the additional wall-clock time for a
workload in which events are distributed evenly across the threads. The table
reports this estimate as a percentage of the total pass time.

With suitable local cache storage, this overhead is only 0.1--1.2\%. In other
words, the first pass takes approximately the direct-read time plus a small
additional cost for filling the cache. Most cache-writing work does not add
directly to the runtime because it occurs while the analysis is already waiting
for data from the remote source. 
The FNAL measurement, at 3.1\%, is worth examining in detail because it shows
when this overhead appears. The two \mFone\ fills wrote the same
143\,GB of pages to the same cache device; the only difference is the rate at
which the source delivered data: 244\,MB/s from CERN against 455\,MB/s from
FNAL. The cache device sustains writes at 564\,MB/s
(Table~\ref{tab:bench}), so the fill used 43\% of its write capability in
the first case and 81\% in the second. Waiting time grows steeply,
not proportionally, as a device approaches saturation. The counters show the
effect directly. From CERN, reading threads stalled 19\,910 times for
5.2\,ms on average; from FNAL, 45\,165 times for 13.9\,ms. That
13.9\,ms is close to the time one coalesced 8\,MiB drain write takes on this device,
so a stalled read was typically waiting out one drain ahead of it. Six
times more stalled thread-time on a pass 1.9 times shorter yields the factor
of 11 between the two percentages. The general rule is that filling is
effectively free while the source delivers data more slowly than the cache
device can absorb it, and the overhead becomes visible as the delivery rate
approaches the device's sustained write rate. The rate-limited cloud volume
below is the same mechanism pushed past that point.

The rate-limited cloud volume behaves differently. Its internal counter reports a
fill overhead of 27.4\%, but the counter measures only the time that reading
threads spend waiting for free space in the staging buffer. On this volume,
cache writes are limited by the available I/O operations. When writes cannot
drain the buffer fast enough, the entire input pipeline slows down: new requests
are issued less frequently, so much of the resulting delay appears as slower
reading from the remote source (34 versus 209\,MB/s in this session) rather
than as time spent waiting for buffer space. The counter also excludes time
spent draining pending writes when files are closed or when the process
finishes. In addition, filling the cache requires more data to be read because
requests are rounded to whole 4\,KiB pages. For these reasons, the 27.4\%
counter value substantially underestimates the actual cost of filling the cache
on this volume. The table therefore uses the measured wall-clock times instead.
The cache-filling pass took 6.1 times as long as direct reading, corresponding
to an overhead of 514.6\%.

The wall-clock comparison of the cold and direct passes is only a
cross-check on these counter-based estimates, not a measurement. Although
the two passes run back to back in one session, the remote source's
delivery rate drifts between them: pairs in these campaigns differ by
6--7\% on average, in both directions, with individual pairs up to about
10\% apart. A comparison at that level cannot resolve a fill cost of a
fraction of a percent. The cross-check is nevertheless consistent with the
counters: cold passes track their direct passes within the source drift
on every suitable configuration. Only on the rate-limited cloud volume,
where the fill costs 6.1 times the direct time and dwarfs any source
drift, is the wall-clock difference itself the measurement.

The same cloud volume also performed poorly after the byte cache had already
been filled: its warm byte tier was 6.9 times slower than direct access, and
even the replica tier ran 3.0 times slower. Only in an earlier campaign, when
the volume delivered more throughput, did the replica tier improve on the
no-cache baseline, by 1.15 times; Sections~\ref{sec:geometry}
and~\ref{sec:limits} examine this volume further. These results show that a
cache is useful only when the storage device can support the I/O pattern of the
workload. A slow or heavily rate-limited device can make caching slower than
reading directly from the remote source.

Even when the cache storage is fast enough, the speedup relative to direct
access depends on the remote source. Two factors are especially important: the
network distance to the source and the load on the source at the time of the
measurement (Section~\ref{sec:origins}). The two \mFone\ rows in
Table~\ref{tab:e2e-lzma} illustrate both effects. The FNAL source has a network
round-trip time of about 26\,ms, compared with 78\,ms for CERN. The direct
access times consequently differ by a factor of 1.6. In contrast, the
warm-cache times differ by only a few percent. Once the data are served from
local storage, analysis time is therefore much less sensitive to network
conditions and load on the remote server.

The workstation provides a useful comparison because it is only 0.30\,ms from
the CERN source. Network latency therefore contributes very little to its
direct-read time. Even in this case, which strongly favors direct access, the
byte cache is 1.6--1.8 times faster. For LZMA-9 data, using replicas provides a
further 1.6 times speedup over the byte cache. Most of this additional gain
comes from reducing the amount of work required for decompression. The CERN
batch node \mFtwo\ is equally close to the source but carries a
current-generation processor, and the same proximity no longer closes the
gap: its byte cache is 3.2--3.8 times faster than direct access and its
replica tier 9.0--11.2 times. Proximity favors direct reading only while the
source can keep up with what the machine consumes.

Overall, repeated analyses run faster with the cache on every tested dataset
and machine configuration that uses sufficiently fast local storage. Building
the cache during the first pass adds little runtime on those configurations.
The absolute warm-cache times are the most stable measurements because they
reproduce closely across runs and do not depend on the current performance of
the remote source. Speedup ratios relative to direct access are still useful
for showing the gains seen in practice, but they are not fixed properties of
the cache. They also depend on network conditions and on the performance of the
remote source at the time of the measurement.

\subsection{File open latency}
\label{sec:openlatency}

End-to-end runtime shows the overall benefit of caching, but not which parts of
the analysis improve or by how much. In this subsection, we examine the
graph-building phase separately. This phase includes opening all input files and
preparing the analysis graph before event processing begins.

A fit to multiple \mFone\ measurements using 787 and 1456 files
across the two datasets gives
\[
\text{remote source: } 28.5\,\text{s} + 4.27\,\text{ms/file}
\qquad
\text{cache: } 8.3\,\text{s} + 1.30\,\text{ms/file}.
\]
Serving files from the cache removes about 20\,s of fixed setup time and
about 3\,ms of overhead per file. Most of the remote-source overhead is
therefore fixed rather than proportional to the number of files. Across
the measured range, growing the file count by 1.85 times increases the fitted
remote graph-building time by only 9\%. These results provide little
motivation to merge files solely to reduce file-opening overhead, although
merging may still benefit other parts of the analysis.

The saving also grows with the distance to the remote source. For the same
dataset, cache serving saves 1.55\,s on a 0.30\,ms local network, about 6\,s
at 17.5\,ms and 27.8\,s at 77.9\,ms. On the same machine, graph building takes
18.3\,s when reading from FNAL at 26\,ms and 39.1\,s when reading from CERN at
77.9\,ms.

These measurements do not identify the exact cause of the fixed cost. Possible
contributors include connection setup, authentication or redirection and schema
discovery during graph construction. The two datasets both contain exactly 9
sample and variation pairs, so this study cannot separate a per-graph
fixed cost from a per-analysis fixed cost. A dataset with a different number of
variations would be needed to distinguish them.

An independent result supports the same interpretation. In the subMIT study,
local-file graph building took 9.0\,s. Our cache-served graph building over
\texttt{root://} URLs took 9.17\,s, while uncached reading of the same URLs in
that study took 45.2\,s~\cite{dalfonso2026submit}.

The conclusion is that the cache improves responsiveness before event processing
begins. It removes most of the delay associated with opening remote files, a
benefit that is especially important for the small, frequently repeated jobs
typical of analysis development. The \ucache plugin also avoids contacting the origin to
check whether cached files have changed while they remain within the freshness
window, which is 7 days by default. This eliminates an additional source of
remote latency when cached files are opened repeatedly.

\subsection{Replica performance gains}
\label{sec:codec}

The replica tier does more than place data nearby. It also converts the data to
a compression format that is faster to decode. This subsection analyzes how much
of the performance gain comes from that conversion and what other factors
contribute.

\begin{table}[t]
\centering\small
\caption{Data and processing characteristics of the two cache representations,
  measured for the two datasets. The upper rows are totals for one complete
  analysis pass over the dataset, and the first row is the volume the
  analysis itself requests, independent of the representation; the lower rows describe the cache on disk: the
  byte cache as a cold pass fills it, the replica store the sweep builds, and
  their combined total once replicas supersede cached bytes. \datasetZ\ was
  measured on the workstation and \datasetL\ on the desktop, both with
  release 0.19.1.}
\label{tab:mechanism}
\begin{tabular}{lrrrr}
\toprule
 & \multicolumn{2}{c}{\datasetZ (zlib-1)} & \multicolumn{2}{c}{\datasetL (LZMA-9)} \\
& byte & replica & byte & replica \\
\midrule
data requested by the analysis (GB) & \multicolumn{2}{c}{92.1} & \multicolumn{2}{c}{120.0} \\
user-mode instructions ($10^{12}$) & 13.1 & 10.0 & 56.1 & 17.5 \\
bytes read off cache (GB)         & 101.9 & 104.5 & 147.0 & 194.6 \\
device read operations ($10^6$)    & 2.38 & 0.17 & 6.58 & 0.36 \\
mean read size (KiB)               & 41.8 & 596 & 21.8 & 535 \\
\midrule
files in cache     & 787 & 787 & 1456 & 1456 \\
size on disk (GB) & 100.5 & 97.1 & 142.9 & 174.1 \\
cache total after reclaim (GB) & \multicolumn{2}{c}{116.4} & \multicolumn{2}{c}{217.8} \\
\bottomrule
\end{tabular}
\end{table}

Table~\ref{tab:mechanism} summarizes the processing characteristics and data
volumes of the byte cache and replica tier for the \datasetZ and \datasetL
datasets. A significant difference can be observed in the number of user-mode
CPU instructions executed by the two tiers. For \datasetZ, the byte cache
requires 1.31 times as many CPU instructions as the replica tier. For
\datasetL, it requires 3.20 times as many, primarily because of data decoding. These differences represent the expected performance gains from
reducing decompression overhead.

The table also quantifies the disk-space requirements of the two caching
strategies. For the zlib dataset, the replicas occupy 97.1\,GB. Because a
replica replaces the cached original bytes from which it was built, the cache
releases the space occupied by those bytes and retains only the original bytes
not covered by replicas, 19.3\,GB in this case. The complete cache therefore
occupies 116.4\,GB, which is 16\% more disk space than caching the original
bytes alone.

For the LZMA dataset (the release-0.19.1 column), the replicas occupy
174.1\,GB, while the complete cache occupies 217.8\,GB, including 43.7\,GB of
retained original bytes. This total is
52\% larger than the space required to cache the original bytes alone. The
difference between the two datasets has a straightforward explanation: ZSTD-1 is
selected for its fast decompression speed, but it achieves a lower compression
ratio than LZMA-9. Consequently, replicas generated from LZMA-compressed data
require more space than the original bytes they replace.






\begin{table}[t]
\centering\scriptsize
\setlength{\tabcolsep}{4pt}
\caption{Byte cache vs recompressed replica performance on different machines
  for two compression algorithms. ``Execution'' is event-loop time: wall-clock
  time minus graph building. ``I/O band,\%'' is the achieved cache-device read
  rate over what uCache's own read path delivers for that tier's shape on the
  same device, at the concurrency listed in Table~\ref{tab:bench}. ``RAM cache'' is
  the ratio of execution times with and without operating-system
  page-cache eviction; for \mM\ it is not available because macOS provides
  no way to evict the page cache selectively. All rows used release 0.19.1,
  except the \mM\ rows (release 0.20.0).}
\label{tab:codec-machines}
\resizebox{\textwidth}{!}{\begin{tabular}{llrcccll}
\toprule
\multirow{2}{*}{Machine} & \multirow{2}{*}{Algorithm} & Execution (s) & \multirow{2}{*}{Speedup} & I/O band & RAM cache & \multicolumn{2}{c}{Limiting resource}\\
                         &                              & byte / replica & & \% & byte / replica & byte & replica \\
\midrule
\mD\ (6t, SSD) & zlib-1 & 309 / 205 & 1.50 & 69 /\;\; 90 & 1.00 / 0.99 & CPU $+$ I/O wait & I/O band \\
\mD\ (12t, SSD) & zlib-1 & 217 / 179 & 1.21 & 99 / 103 & 1.00 / 1.00 & CPU $+$ I/O band & I/O band \\
\mD\ (6t, NVMe) & zlib-1 & 217 / 127 & 1.71 & 29 /\;\; 23 & 1.06 / 1.00 & CPU & CPU \\
\mD\ (12t, NVMe) & zlib-1 & 145 / 101 & 1.44 & 43 /\;\; 30 & 0.99 / 1.00 & CPU & CPU \\
\mW\ (32t) & zlib-1 & 212 / 181 & 1.17 & 88 / 103 & 1.86 / 2.24 & I/O band & I/O band\\
\mFone\ (64t) & zlib-1 & 40 /\;\; 25 & 1.61 & 91 /\;\; 86 & 1.29 / 1.00 & CPU $+$ I/O wait & CPU \\
\mFtwo\ (64t) & zlib-1 & 60 /\;\; 20 & 2.94 & 67 /\;\; 80 & 2.36 / 1.16 & I/O wait & CPU $+$ I/O wait \\
\mV\ (24t, NVMe) & zlib-1 & 116 /\;\; 71 & 1.62 & 40 /\;\; 15 & 1.01 / 1.00 & CPU & CPU \\
\mV\ (24t, cloud) & zlib-1 & 3299 / 1427 & 2.31 & 77 /\;\; 39 & 1.03 / 1.01 & I/O op cap & I/O band \\
\midrule
\mD\ (6t, SSD) & LZMA-9 & 1057 / 355 & 2.98 & 29 /\;\; 98 & 1.00 / 1.01 & decode CPU & I/O band \\
\mD\ (12t, SSD) & LZMA-9 & 684 / 319 & 2.15 & 45 / 109 & 0.97 / 1.01 & decode CPU & I/O band \\
\mD\ (6t, NVMe) & LZMA-9 & 1001 / 208 & 4.82 & 9 /\;\; 27 & 1.01 / 0.99 & decode CPU & CPU \\
\mD\ (12t, NVMe) & LZMA-9 & 639 / 169 & 3.79 & 14 /\;\; 33 & 1.01 / 1.00 & decode CPU & CPU \\
\mW\ (32t) & LZMA-9 & 524 / 320 & 1.64 & 52 / 109 & 1.04 / 1.46 & decode CPU & I/O band \\
\mFone\ (64t) & LZMA-9 & 138 /\;\; 43 & 3.18 & 39 /\;\; 93 & 1.07 / 1.05 & decode CPU & CPU \\
\mFtwo\ (64t) & LZMA-9 & 139 /\;\; 33 & 4.20 & 41 /\;\; 92 & 1.16 / 1.06 & decode CPU & CPU \\
\mM\ (6t) & LZMA-9 & 1434 / 546 & 2.63 & \;\;9 /\;\;\;\; 9 & -- & decode CPU & CPU \\
\mM\ (10t) & LZMA-9 & 1098 / 441 & 2.49 & \;\;7 /\;\; 12 & -- & decode CPU & CPU \\
\bottomrule
\end{tabular}}
%
\end{table}

Table~\ref{tab:codec-machines} shows what the same two representations
deliver on the test machines, with event-loop time separated from graph
building. If removing decompression instructions were the whole story,
the speedup would be the instruction ratio of
Table~\ref{tab:mechanism} (1.31 for zlib-1 and 3.20 for LZMA-9 at full
replica coverage) on every machine. The deviations identify each machine's limit. The
workstation falls short because its SATA device is the limiting resource: its replica
passes read at 102 to 109\% of what the cache's read path delivers on
that device, and evicting the page cache roughly doubles its zlib-1
times. The
desktop, running 6 threads on 6 cores, falls slightly short of the
full-coverage expectation on its SATA device (2.98 against 3.20) and
exceeds it on its NVMe device (4.82 against 3.20): the SATA replica pass
runs at 98\% of the path's rate and waits on I/O, which the instruction
ratio does not account for, while the NVMe replica pass (27\% of its rate)
does not. The desktop's zlib-1 speedup at 6
threads on SATA exceeds the expectation (1.50 against 1.31) for the same
reason on the byte side: the byte pass makes
2.4 million small reads on its SATA device (served at 0.48\,GB/s
through the cache's read path)
and waits on I/O, which instruction counts do not capture, while the
replica pass reads at 90\% of its path's rate. At 12 threads the
replica pass itself reaches the path's full rate (103\%), waits the same
way, and the speedup falls below the ratio (1.21). On the NVMe device neither
tier approaches its path's rate and the speedup, as for LZMA-9, exceeds
the instruction ratio (1.71 at 6 threads, 1.44 at 12). Its equal
evicted and RAM-resident times do not rule the device out: with
16\,GB of RAM the cached data never fit in the page cache, so both
passes read from the device. \mFone's
zlib-1 speedup exceeds the expectation for the same reason as the
desktop's. Its
LZMA speedup of 3.18 against the expected 3.20 is the clean case:
nothing else limits the run, and the replica delivers almost exactly the
instruction reduction. The CERN batch node \mFtwo\ shows the byte tier's
per-request cost most clearly. At 64 threads its byte passes run at 41 and
67\% of what the cache's read path delivers on this device, and that path
itself reaches only 47\% of the raw hardware for this scattered shape
(it achieves a queue depth of 12 while 64 threads are issuing requests), yet the
same passes run 1.16 to 2.36 times faster when served from memory: the
cost is waiting on individual requests, not bandwidth, with 2.4 to 6.6
million reads, each with a latency that instruction counts do not
capture.
The replica's few large reads remove both gaps at once: its path
matches the raw device and its passes reach 80 and 92\% of it, which
is why both of \mFtwo's speedups exceed their instruction ratios (2.94
against 1.31 and 4.20 against 3.20).

The benefit has a storage cost. As discussed in Section~\ref{sec:opscost},
replicas of the LZMA-9 dataset increase the total cache footprint to 1.52 times
the byte-only footprint. For the zlib-1 dataset, the total is 1.16 times the
byte-only footprint.

The conclusion is that the replica tier is most valuable when decompression is a
major part of the analysis cost. Its speedup is determined by the source
compression format and the processor, not by locality alone. The trade-off is a
larger cache footprint, especially when the source uses a dense but expensive
format such as LZMA-9.

\subsection{Storage access pattern and the limiting resource}
\label{sec:geometry}

The previous subsection showed that reducing decompression cost can have a large
effect on performance when decompression is the limiting resource. This
subsection considers a different question: how does the shape of storage
requests affect performance, and under what conditions does reducing the number
of requests actually make the analysis faster?

The two tiers request the same data, but they organize those requests very
differently (Table~\ref{tab:mechanism}). The byte tier issues approximately one
device read for each logical request. Across both datasets, it performs 1.002 to
1.004 operations per request, which is close to the minimum possible. As a
result, the device reads closely follow the access pattern generated by the
analysis itself. Each pass therefore consists of millions of operations, with an
average read size of 21.8 to 41.8\,KiB.

The replica layout changes this request shape substantially. It delivers the
same data using 14 to 18 times fewer reads, and each read is 14 to 25 times
larger. However, this reduction in operation count improves performance only if
the storage device is limited by the rate at which it can process requests. On
the local NVMe and SATA devices used in this study, neither tier comes close to
the device request-rate limit. The warm-run times are instead determined by
decompression cost or storage bandwidth, as discussed in
Sections~\ref{sec:endtoend} and~\ref{sec:codec}. Under those conditions,
reducing the number of reads does not remove the resource that limits the run,
so it provides little or no performance benefit.

A deliberately constrained cloud block volume illustrates the opposite case,
in which request shape does determine performance. Its device test delivers
about 1000 IOPS regardless of
concurrency, compared with roughly $10^5$ on the machine's local disk. On
this volume the byte tier was 6.9 times slower than using no cache: it
issued 2.38 million operations, the volume completed them at about 720 per
second, and the wall-clock time is the operation count divided by that rate.
Reducing the operation count is the only change that can help.

The replica tier is that change: it delivered the same data in 14.4 times
fewer operations of 14.7 times the size (613\,KiB reads), which cut the
penalty by 2.3 times, yet the pass still ran 3.0 times slower than direct
reading. An earlier campaign on the same volume had measured the identical
replica pattern (the same bytes, operation count and request size) at
2.9 times the throughput, fast enough to exceed direct reading. The delivered
rate of a shared volume with an operation quota is therefore itself a variable
of the environment, and no cache tier can guarantee a stable result on such a
device. What is
stable across both campaigns is the mechanism: fewer, larger requests are
strictly better on a device limited by its operation rate.

This example shows why reducing the number of storage operations does not
automatically reduce total analysis time. A change improves performance only
when it reduces pressure on the resource that limits the run. If
decompression is the limiting resource, reducing storage operations does not
address the bottleneck. If bandwidth is the limiting resource, larger requests
help only if they change the achieved bandwidth. If the device is limited by
IOPS, however, reducing the number of operations can directly reduce execution
time. Table~\ref{tab:regimes} summarizes the 4 limiting resources observed in
this study and the changes that improved performance in each case.

\begin{table}[t]
\centering\small
\caption{The resource that limits each measured case, the evidence used to identify it and the change that improved performance.}
\label{tab:regimes}
\begin{tabular}{>{\raggedright\arraybackslash}p{0.15\textwidth}>{\raggedright\arraybackslash}p{0.29\textwidth}>{\raggedright\arraybackslash}p{0.27\textwidth}>{\raggedright\arraybackslash}p{0.19\textwidth}}
\toprule
Limiting resource & Evidence & Measured example & Effective response \\
\midrule
remote-source latency & few CPU cores are busy; direct and cold runs keep improving as more streams are added & all direct and cold runs on links with 17.5 to 78\,ms latency & use a local cache or more concurrent streams \\
storage request rate & requests per second approach the tested device limit; wall-clock time follows request count & byte tier on a rate-limited cloud volume: wall-clock time equals request count divided by delivered rate & use the replica layout ($14.7\times$ more data per request) \\
storage bandwidth & data rate approaches the tested device limit after reads have been combined & \datasetZ byte tier on SATA: 89\% of the 0.545\,GB/s the cache's read path delivers there & use a faster device or reduce the stored data \\
decompression CPU & storage request rate and bandwidth remain below their limits; profiles are dominated by decompression & \datasetL byte tier: 39\% of \mFone's through-cache read rate; 51\% of SATA bandwidth on the workstation, IPC 1.41 & recompress the data ($1.6$ to $4.8\times$) \\
\bottomrule
\end{tabular}
\end{table}

The practical conclusion is that the limiting resource should be identified
before changing the cache or storage layout. Different bottlenecks require
different remedies: remote latency, device operation rate, device bandwidth, and
decompression cost each respond to different optimizations. In the measurements
reported here, the plugin counters together with the \texttt{ucache bench}
device test provide enough information to determine which of these resources is
limiting performance.

\subsection{Remote data source variability}
\label{sec:origins}

Any speedup measured relative to direct remote reading depends on both sides of
the comparison: the performance of the cache and the performance of the remote
data source. The remote source is not necessarily stable, especially when it is
shared production infrastructure. This subsection measures how much source
performance can vary and explains why absolute warm-cache times are a more
reliable performance measure than cache-to-direct ratios.

Four observations were especially important:

\begin{enumerate}
\item \textbf{The performance of a shared source can change quickly.} Two fills
  of the same dataset from the same host, separated by 51 minutes, took 634\,s
  and 1298\,s, a factor of 2.05 difference. Throughput from CERN to one client
  repeatedly varied from 52 to 211\,MB/s within an hour. This level of
  variability invalidated 3 experiments that compared remote-source runs
  performed at different times because changes in the source could not be
  separated from changes caused by the system under test. Warm-cache
  performance was much more stable. Warm-replica runs took 199.93, 199.95,
  198.05 and 199.58\,s across 3 software versions and 3 days, a spread
  of only 0.96\%.

\item \textbf{A direct site endpoint can be much faster than a federation
  redirector.} A federation redirector chooses among many data servers, so a
  single analysis can receive files from several sites with different network
  paths and storage conditions. Reading \datasetL through the CMS US regional
  redirector, \texttt{cmsxrootd.fnal.gov}, distributed 1456 files across 63
  servers at 10 sites on 2 continents. Round-trip times ranged from 6.8 to
  150\,ms, and aggregate throughput was 42 to 54\,MB/s. Reading the same
  dataset through FNAL's direct site endpoint,
  \texttt{cmsdcadisk.fnal.gov}, sustained 407\,MB/s. Thus, changing only the
  hostname increased throughput by a factor of 8 and changed the job from
  I/O-limited to compute-limited. A single-file test did not expose
  this difference because it selected one nearby replica. The problem became
  visible only when many files were read concurrently and the redirector
  distributed those files across many servers.

\item \textbf{Network distance alone does not predict throughput.} The MIT
  Tier-2 endpoint was 6.9\,ms from the client but sustained only 29 to 53\,MB/s
  and slowed during the run. CERN was 77.9\,ms away and sustained about
  110\,MB/s. FNAL was 26\,ms away and sustained 407\,MB/s. These measurements
  show that a shorter round-trip time does not necessarily produce higher
  throughput. In these runs, load on shared production storage had a larger
  effect than network distance.

\item \textbf{The client network configuration can make distributed reads
  worse.} With the default Linux setting
  \texttt{net.ipv4.tcp\_slow\_start\_after\_idle=1}, bursty XRootD traffic
  repeatedly re-entered TCP slow start after pauses for decompression. Of 381
  live streams, 357 had a congestion window of 10 packets, or about 14\,KB in
  flight, while the best stream reached 1318\,Mbps. Distributing files across
  63 endpoints reduced the amount of sustained traffic on each individual
  connection. As a result, many connections did not remain active long enough
  to grow their sending windows before another idle period caused slow start to
  begin again.
\end{enumerate}

The cross-source control makes the effect of source variability especially
clear. Between the CERN and FNAL full-benchmark runs, the direct
remote-reading time changed by a factor of 1.6. Over the same comparison, the
warm-cache times changed by at most 2\%. The cache measurements were therefore
far more stable than the remote-reading baseline used to compute speedup ratios.

The practical consequence is that direct-reading performance is an unstable
baseline on shared wide-area infrastructure. Cache performance should therefore
be reported primarily using absolute warm-cache times and repeated
measurements. Ratios relative to direct reading remain useful because they
reflect what happens in practice, but they should always identify the source,
endpoint, and measurement conditions so that changes in the remote system are
not mistaken for changes in cache performance.

\subsection{Costs and break-even point}
\label{sec:opscost}

A cache is useful only if its performance benefit is large enough to justify
the costs of creating and maintaining it. These costs include the overhead of
the first pass, the additional disk space, and the background work required to
build replicas. This subsection measures each of these costs and determines how
many analysis passes are required before the replica tier recovers its
construction cost.

\paragraph{First-pass overhead.}
The cache is designed to fill during the first useful analysis pass rather than
requiring a separate preprocessing step. The design target is for this first
pass to take no more than 10\% longer than the direct-read time. The fill
overhead measured by the plugin's own counters (Section~\ref{sec:endtoend}) is
0.1--1.2\% of the pass on every configuration with suitable local storage, and
3.1\% against the fastest source, well inside the target. Cold-versus-direct
wall-clock comparisons give consistent results, but they cannot reliably resolve
an effect this small because the remote source itself varies by several percent
between individual runs. The exception is the rate-limited cloud volume, where
filling the cache cost 6.1 times as much as a direct read
(Section~\ref{sec:limits}).

\paragraph{Disk footprint.}
After one complete pass and a replica-building sweep, \datasetZ uses 116.4\,GB
of disk space: 19.3\,GB for cached source bytes and 97.1\,GB for replicas.
This is 1.16 times the byte-only footprint. \datasetL uses 217.8\,GB:
43.7\,GB for cached source bytes and 174.1\,GB for replicas. This is 1.52
times the byte-only footprint because ZSTD-1 replicas are larger than the
original LZMA-9 bytes. Superseded byte pages are removed from sparse cache files
so that the total footprint remains bounded. A stricter setting,
\texttt{recompress\_reclaim = full}, drops the entire original byte copy once
a file's replica has been validated, so the cache shrinks to approximately the
replica store alone: 97.1\,GB for \datasetZ\ and 174.1\,GB for \datasetL.
The saved space has a cost: reads that replicas do not cover, including those
made when a file is opened, are fetched from the remote source again,
measured at about 12\,KB per file on the first pass after the reclaim, or
0.01\% of the served bytes.
The default setting keeps those bytes local, so warm passes never contact the
remote source while the cache holds the working set. In both cases, the cache stores only 6 to 9\% of the full dataset, rather
than storing the entire dataset.

\begin{table}[t]
\centering\small
\caption{Reaching a fully replicated cache by building replicas during the
  first pass and finishing with one sweep. \emph{Building overhead} is the
  additional time reading threads spent waiting on the staging buffer
  compared with a plain fill in the same session, measured by the same
  counters as the fill-overhead column of
  Tables~\ref{tab:e2e-lzma} and~\ref{tab:e2e-zlib}. \emph{Replicas at exit}
  is the fraction built when the analysis finished. \emph{Total / plain
  fill} divides the total (fill plus finishing sweep) by the estimated plain
  fill under the same conditions (the fill time minus the overhead).
  \emph{Background workers} is the default worker count, computed from the
  rule above rather than recorded.
  The Mac mini and \mFtwo\ campaigns built replicas with explicit sweeps only
  and are not shown. All rows used release 0.19.1.}
\label{tab:bgbuild}
\begin{tabular}{lrrrrrr}
\toprule
 & \makecell{Backgr.\\workers} & \makecell{Fill with\\building (s)} & \makecell{Building\\overhead (s)} & \makecell{Replicas\\at exit} & \makecell{Finishing\\sweep (s)} & \makecell{Total /\\plain fill} \\
\midrule
\multicolumn{7}{l}{\datasetZ (zlib-1)} \\
\mD\ (6t, SSD)                & 3  & 1969 & 0.5  & 98.9\% & 6    & 1.00 \\
\mD\ (6t, NVMe)               & 3  & 1955 & 0.3  & 98.6\% & 6    & 1.00 \\
\mD\ (12t, SSD)               & 3  & 1381 & 0.2  & 98.3\% & 5    & 1.00 \\
\mD\ (12t, NVMe)              & 3  & 1357 & 0.2  & 98.6\% & 6    & 1.00 \\
\mV\ (24t, NVMe)              & 12 & 477  & 0.2  & 95.2\% & 22   & 1.05 \\
\mW\ (32t)                    & 16 & 615  & 85.9 & 40.8\% & 375  & 1.87 \\
\mFone\ (64t)                    & 32 & 394  & 18.7 & 93.1\% & 13   & 1.08 \\
\midrule
\multicolumn{7}{l}{\datasetL (LZMA-9)} \\
\mD\ (6t, SSD)                & 3  & 3553 & 0.8  & 99.5\% & 15   & 1.00 \\
\mD\ (6t, NVMe)               & 3  & 3539 & 0.9  & 99.3\% & 15   & 1.00 \\
\mD\ (12t, NVMe)              & 3  & 2171 & 0.8  & 98.6\% & 14   & 1.01 \\
\mW\ (32t)                    & 16 & 1054 & 64.2 & 35.5\% & 1215 & 2.29 \\
\mFone\ (64t)                    & 32 & 619  & 25.1 & 97.1\% & 28   & 1.09 \\
\bottomrule
\end{tabular}
\end{table}

\paragraph{Background replica building.}
With \texttt{recompress = on}, replicas are built in the background while the
first pass is still reading data. By default, the number of background workers
is half the machine's core count, or half the process's thread count if that
number is smaller. For a user who intends to use the replica tier, three
practical questions matter: how much of the replica store has been built when
the first pass finishes, how much the background work slows that pass, and how
much additional time is required to reach a fully replicated cache compared
with filling the byte cache alone. Table~\ref{tab:bgbuild} answers all three
questions for each configuration. It reports the most time-efficient route to
a fully replicated cache: build replicas during the fill, then complete any
remaining replicas with one explicit sweep.

On every machine where the fill leaves processing capacity idle, either because
the desktop is reading from a distant source or because the cloud VM and
\mFone\ have spare cores, background replica building adds little cost
and is nearly complete when the analysis finishes. The measured blocking
overhead is below one second on the desktop and the VM, and 19 to 25\,s
($\sim$4--5\%) on \mFone. By the end of the fill, 90 to 99\% of the
replicas have already been built. The remaining sweep therefore takes only
seconds, and the total time to reach a fully replicated cache is 1.00 to
1.09 times the time required for a plain fill.

The workstation is the limiting case. It runs 32 analysis threads on 32 cores
against a nearby source, leaving essentially no idle processing capacity for
background construction. The measured blocking overhead rises to 64 to 86\,s,
but this accounts for only about half of the actual slowdown. Wall-clock
comparisons with plain fills performed in the same sessions show that fills
with background building are 21 to 47\% slower across 3 campaigns. The
remaining slowdown comes from background workers competing with analysis
threads for CPU time without causing explicit thread blocking. This is the same
limitation of the blocking counter discussed for the cloud volume in
Section~\ref{sec:endtoend}. Consequently, only 35 to 41\% of the replicas have
been built when the first pass finishes. The remaining sweep is correspondingly
long, and the total time to reach a fully replicated cache is 1.9 to 2.3 times
the time required for a plain fill.

Background replica building is therefore most effective when the first pass
leaves compute capacity unused. In that regime, most replica construction can
be hidden behind the fill, so a fully replicated cache is available with little
additional wall-clock cost. On a host where the analysis occupies every core and the source keeps them
busy, background workers instead compete directly with the analysis, and building during the fill provides no
advantage over filling the byte cache first and then running one replica sweep.

\paragraph{Break-even point.}
The byte cache recovers its cost on the first repeated pass on every suitable
configuration: filling costs 0.1--3.1\% of the first pass
(Section~\ref{sec:endtoend}), and every later pass is 1.6 to 8.8 times faster
than reading remotely (Tables~\ref{tab:e2e-lzma} and~\ref{tab:e2e-zlib}). The
question that requires arithmetic is the replica tier. Its extra cost is the
building overhead plus the finishing sweep (Table~\ref{tab:bgbuild}); its
return is the difference between a byte-tier and a replica-tier pass.

Where the fill has idle capacity, the investment returns within the first
repeated pass. On the desktop (12 threads, NVMe, \datasetL), the replica tier
costs about 15\,s of overhead and finishing sweep, and saves 471\,s on every
later pass. On \mFone\ it costs about 53\,s and saves 96\,s per
pass. On \mFtwo, which built replicas with one explicit sweep after the fill,
the sweep costs 42 to 118\,s against savings of 40 to 107\,s per pass, so
the investment returns after about one repeated pass. The workstation is again the limiting case: the cost is about
1400\,s (a fill slower by 190\,s and a 1215\,s finishing sweep) against a
saving of 205\,s per pass, so it breaks even after roughly 7 repeated
passes. For \datasetZ, where a replica pass saves only 31\,s over the byte
tier, about 20 passes are needed. This multi-pass break-even arises only in the workstation's combination of
circumstances: the analysis occupies every core, and the nearby source is
fast enough to keep them busy. With a more distant or slower source, even a
machine running analysis threads on all of its cores has idle cycles during
the fill (the desktop, with 6 threads on 6 cores, is such a case),
and the replica tier recovers its cost immediately, and the benefit grows
with every additional pass over the same data.

\begin{table}[t]
\centering\small
\caption{The \xcache configurations used in the measurements: the default
  configuration and the configuration tuned for the AGC benchmark. Both used
  the same disk-capacity limits (\texttt{pfc.diskusage}).}
\label{tab:xcache-config}
\begin{tabular}{lllp{8cm}}
\toprule
Parameter & Default & Tuned & Impact \\
\midrule
\texttt{pfc.prefetch} & 10 & 0 & Controls the in-flight depth of the
  whole-file prefetcher. Any nonzero value causes complete files to be cached;
  0 fetches only the requested blocks and keeps the cache footprint bounded. \\
\texttt{pfc.ram} & 1g & 8g & Sets the RAM available for blocks in flight.
  At the default value, blocks can overflow to an uncached pass-through path. \\
\texttt{oss.preread} & off &  \makecell[tl]{32 limit 16m\\qsize 256} & Enables server-side
  read-ahead when cached blocks are served from disk; in these measurements,
  it roughly halves the warm-pass time. \\
\texttt{pfc.blocksize} & 128k & 128k & Sets the cache block size and was
  left at its default value. A 16-fold sweep changed the warm-pass time by
  only 1.8\%. \\
\bottomrule
\end{tabular}
\end{table}

\begin{table}[t]
\centering\scriptsize
\caption{Results from \ucache and \xcache on the full AGC dataset, using the
  same SSD and 32 analysis threads. The no-cache baseline was 384.66\,s,
  measured in a separate session. The fill's $\approx$1$\times$ direct is
  supported by back-to-back cold and direct passes in one session (0.995 and
  0.999 times); the raw 465/384.66 ratio compares across sessions on a path
  whose direct time drifts by about 9\%. The \ucache run used
  \texttt{recompress\_reclaim = full}; with the default reclaim its footprint is
  116.4\,GB. The default-configuration \xcache footprint was capped by its disk
  high-watermark; its demand was about 1.6\,TB. Replicas covered all
  787 files.}
\label{tab:xcache_local}
\resizebox{\textwidth}{!}{\begin{tabular}{lrrr}
\toprule
 & \ucache & \xcache default & \xcache tuned \\
\midrule
cold cache (fill)     & 465\,s ($\approx$1$\times$ direct) & 5770\,s (15.0$\times$ direct) & 8698\,s (22.6$\times$ direct) \\
warm cache, byte  & 230\,s & 5092\,s (22.1$\times$ \ucache) & 238\,s (1.03$\times$ \ucache) \\
warm cache, replica  & 201\,s & -- & -- \\
cache footprint            & 99.0\,GB (5.7\%) & 806--876\,GB & 311.1\,GB \\
remote bytes (fill)        & 100.6\,GB & 1563\,GB & 311.1\,GB \\
remote operations (fill)   & $7.3\times10^{3}$ & $11.9\times10^{6}$ & $2.37\times10^{6}$ \\
\bottomrule
\end{tabular}}
\end{table}

\begin{table}[t]
\centering\small
\caption{Results of \ucache on the analysis machine against a tuned \xcache one
  10\,GbE / 0.30\,ms hop away: full AGC dataset, 24-core EPYC client,
  16 analysis threads. The no-cache baseline was 600.3\,s. The \ucache run
  used the default reclaim setting. The \xcache warm
  wall is the mean of two passes that agree to 0.05\%. Replica coverage was
  784 of 787 files (99.6\%).}
\label{tab:xcache_remote}
\begin{tabular}{lrr}
\toprule
 & \ucache\ (client NVMe) & \xcache\ tuned (one hop) \\
\midrule
cold (fill)                & 588.6\,s (0.98$\times$ direct) & 8372.9\,s (13.9$\times$ direct) \\
warm, byte tier            & 163.0\,s & 245.5\,s \\
warm, replica tier         & 100.7\,s & -- \\
cache footprint            & 116.4\,GB (6.5\%) & 311.1\,GB \\
remote bytes (fill)        & 100.5\,GB & 311.1\,GB \\
remote operations (fill)   & $6.8\times10^{3}$ & $2.37\times10^{6}$ \\
\bottomrule
\end{tabular}
\end{table}

\subsection{Comparison with \xcache}
\label{sec:xcache}

\xcache and \ucache address related but different use cases. \xcache, implemented
by \texttt{XrdPfc}~\cite{bauerdick2014xcache}, is a shared proxy cache typically
operated as part of a site's infrastructure. The \ucache plugin is instead used directly
by an individual physicist and runs on the analysis machine. Despite these
differences, comparing their performance is useful because both can serve the
same underlying physics-analysis workload. We therefore evaluate how they
perform for an analysis such as AGC, where repeated passes read only a subset
of a much larger remote dataset.

We compare the performance of \ucache, \xcache and direct remote reading in 2
experimental setups. The first places both caches on the workstation and uses the
same SSD for cache storage. This setup isolates differences between the cache
implementations and evaluates \xcache as a personal local cache.

The second setup uses 2 machines and represents a more typical facility
deployment. \xcache runs on the workstation, while the analysis and \ucache run on
the VM. The analysis therefore accesses \xcache over the local network but
accesses \ucache on the same machine. The network between the two machines is
a single 10\,GbE hop with 0.30\,ms round-trip time, measured to sustain
1.17\,GB/s (94\% of line rate). This experiment tests whether an \xcache
deployed on a nearby cache server can provide performance comparable to a
locally installed \ucache. The setup is favorable to \xcache because the
\xcache instance on the workstation is dedicated to a single analysis and is not
shared with other users, avoiding contention that could reduce cache
performance under production load.

We tested \xcache 5.9.6 with 2 configurations: the default configuration and
a configuration tuned for this workload. Table~\ref{tab:xcache-config}
summarizes the parameters that differ between them and the effect of each
change. The tuned configuration is the primary basis for comparison with
\ucache because it adapts \xcache to the sparse access pattern of the analysis.
We also report results for the default configuration to show the performance
obtained without workload-specific parameter tuning and to quantify how
strongly the \xcache results depend on configuration. By contrast, \ucache
ran with its shipped defaults in every run of every campaign in this paper:
no parameter search was performed for it, and its read path is deliberately
knob-free, so there is little to sweep. The only non-default settings enable
features (the replica tier, the list of source codecs eligible for it, and,
on two campaigns, the space-reclaim mode \texttt{recompress\_reclaim = full})
rather than tune performance. The tuning asymmetry in this comparison
therefore favors \xcache.

For these benchmarks, we used the full AGC \datasetZ dataset. The dataset
contains 787 files and occupies 1.78\,TB at the origin, while the analysis
reads only about 92\,GB during each pass (\ucache fetches 101\,GB of it,
after rounding the requests to whole cached pages). Thus, each pass accesses only a
small fraction of the stored data. This sparse access pattern is important for
the comparison because \ucache and \xcache differ in how much additional data
they fetch and retain while satisfying those requests.

Filling \ucache adds negligible overhead to the first analysis pass, as
established in Section~\ref{sec:endtoend}. The first-pass comparison with
\xcache therefore measures an important practical difference between the two
caches: whether populating the cache substantially delays the first useful
analysis pass. The warm-pass measurements then compare their performance after
the requested data have been cached.

Tables~\ref{tab:xcache_local} and~\ref{tab:xcache_remote} summarize the
measurements for the local and remote \xcache configurations, respectively.
Together, these results highlight five important differences between \xcache
and \ucache:

\begin{itemize}
\item \textbf{Warm byte serving is similar when both caches use the same
  machine and storage.} Tuned \xcache completes the repeat pass in 237.6\,s,
  while the \ucache byte tier completes it in 229.7\,s, a difference of
  3.4\%. We therefore treat their warm byte-serving performance as effectively
  equal in this configuration. The additional warm-path benefit of \ucache
  comes from the replica tier. Replica serving is 1.18 times faster than tuned
  \xcache for the full AGC dataset and 1.32 times faster for a smaller LZMA
  dataset.

\item \textbf{The larger differences are in first-pass cost and disk use.}
  The \ucache fill runs at approximately direct-reading speed, and the cache
  stores only 5.7\% of the dataset. Tuned \xcache takes 22.6 times the direct-reading time to fill
  and uses 3.1 times as much disk as \ucache. The difference is especially
  important for sparse analyses because the useful working set is much smaller
  than the complete dataset. With the default configuration, \xcache fetches
  about 88\% of the full dataset even though the analysis reads only a small
  fraction of it.

\item \textbf{The default \xcache configuration performs poorly when the
  requested footprint exceeds the available cache space.} In this experiment,
  the cache was 1.68 times smaller than the requested footprint, so \xcache
  repeatedly purged complete files. With \texttt{prefetch 10}, a later miss
  causes the complete file to be fetched again. The repeat pass consequently
  issues 99\% as many remote operations as the first pass and transfers about
  1.58\,TB on every pass. Instead of accelerating the analysis, the warm
  \xcache pass is 13.2 times slower than direct reading.

  Setting \texttt{prefetch 0} changes this behavior by fetching only requested
  blocks. The working set then fits within the available cache space, and the
  warm-pass time improves by a factor of 21.5 relative to the default
  configuration. This change has a cost: the first pass becomes 1.5 times
  slower. Changing the cache block size has little effect by comparison. A
  16-fold sweep of \texttt{pfc.blocksize} changes the warm-pass time by only
  1.8\%.

\item \textbf{Much of the \xcache first-pass cost comes from the current
  caching path rather than from block caching itself.} Part of the
  fill-time difference follows directly from design goals: a whole-file
  prefetcher is expected to move more data, and the default configuration's
  fill moves 15.5 times \ucache's bytes in 12.4 times \ucache's fill time:
  volume explains it entirely. The tuned configuration does not have
  that explanation: it moves 3.1 times the bytes in 18.7 times the time,
  and the residual factor of 6 is the caching path itself. In the caching
  path,
  \texttt{XrdPfc} issues one \texttt{Read()} for each 128\,KiB block in a
  serial loop in \texttt{File::ProcessBlockRequest}. Its pass-through path can
  already combine as many as 1024 chunks into one vector read, but this
  batching is not used while the data are being cached. Both caches must
  satisfy about 2.37 million chunks with a mean requested size of 41.4\,KiB.
  The \ucache plugin fetches those chunks in about 7000 vector reads, whereas \xcache
  issues 2.37 million block reads. The byte amplification of 3.09 follows
  directly from $128/41.4$.

  A matched-depth microbenchmark attributes about a factor of 2.6 of the
  first-pass difference to batching. The remainder is not explained by the
  measurements. Across all tested scales, the \xcache fill sustained only about
  273 remote operations per second, and fill time increased linearly with
  dataset size.

  The source review also identified a related configuration limitation,
  with the caveat that it matters for the workload measured here rather than
  for \xcache's primary one: whole-file prefetching is a defensible default
  for a shared site cache, where a fetched file is amortized across many
  users, and it becomes a cost specifically in the sparse single-user
  regime. \texttt{XrdPfc}
  uses prefetch depth as its control over sparse fetching, while the prefetch
  target is the complete file. With the current implementation, it cannot be
  configured both to prefetch sparse requests aggressively and to keep the
  retained cache footprint bounded. Separating the fetch unit from the
  cache-retention unit would remove this coupling. During this work, we also
  found an out-of-bounds read in \texttt{XrdOssFile::ReadV} under
  \texttt{oss.preread}, which we have reported to the \xcache developers
  together with a one-line fix.

\item \textbf{A nearby dedicated \xcache server remains slower than a local
  \ucache for this workload.} Table~\ref{tab:xcache_remote} shows the results
  for the two-machine configuration. The VM runs the 16-thread analysis and
  \ucache, while \xcache runs on the workstation across a 10\,GbE link with
  0.3\,ms latency. Both configurations use the same remote source and dataset.

  In this setup, \ucache fills 14.2 times faster, issues 349 times fewer remote
  operations and requires only 37\% of the disk space of tuned \xcache. On repeat
  passes, the \ucache byte tier is 1.51 times faster than \xcache, and the
  replica tier is 2.44 times faster. Relative to direct remote reading, the
  speedup is 5.96 times for \ucache replica serving and 2.45 times for
  \xcache.

  The local network does not limit either warm run. \xcache and \ucache deliver
  395 and 595\,MB/s, respectively, while the link sustains 1172\,MB/s. The
  difference is instead consistent with the storage available to each cache.
  The VM's local NVMe provides 2.8 times the 4\,KiB operation rate and 3.6
  times the bandwidth of the \xcache storage path. Memory caching does not
  explain the result: the dataset is larger than memory, and runs with data
  resident in the operating system page cache take 0.99 to 1.01 times as long
  as runs after eviction. The observed difference therefore reflects the
  storage paths used by the two caches rather than the network link or the
  operating system page cache.
\end{itemize}

In conclusion, for this sparse AGC workload, \ucache's advantages over
\xcache are concentrated where the two designs differ: first-pass cost, disk
footprint, and the replica tier. When both caches run on the workstation and use
the same SSD, tuned \xcache and the \ucache byte tier provide similar warm
byte-serving performance, but \ucache reaches that state with negligible
first-pass overhead and substantially lower disk use. Its replica tier then
reduces repeat-pass time further. In the two-machine configuration, where
\xcache runs on a dedicated nearby server, the \ucache byte tier is
1.51 times faster and the replica tier is 2.44 times faster. These measurements
show that, for an individual physicist repeatedly processing a sparse subset of a
large remote dataset, \ucache provides both a much lower cost to populate the
cache and better subsequent analysis performance.

\subsection{RNTuple}
\label{sec:rntuple}

This subsection asks whether the caching gains measured so far depend on the
container format. It also separates two sources of improvement in ROOT's
RNTuple format: a more efficient storage layout and a cheaper compression
format. Finally, it measures the limiting case in which neither the layout
nor the compression format leaves anything for the cache to remove.

RNTuple~\cite{blomer2020rntuple,lopezgomez2022rntuple} is ROOT's newer
columnar format. Compared with the TTree layout, its larger pages and grouped
cluster reads reduce the number of storage operations, although the data are
still organized by clusters rather than entirely by branches. These
improvements do not eliminate two considerations that matter for analysis
caching. First, most existing data are stored as TTrees, and physicists reading
published datasets generally do not choose the container format. Second,
decompression cost still depends on the compression format. CMS NanoAOD is
stored with LZMA, so converting frequently read data to ZSTD-1 can remain useful
even when the underlying container is RNTuple.

To isolate the effect of the container, the measurement holds the logical data,
compression format, source and analysis constant. A 1453-file copy of the
\datasetL dataset was made on a separate EOS instance at CERN. It is a subset
of the 1456-file dataset stored on the \texttt{eospublic} EOS instance
(Table~\ref{tab:e2e-lzma}). The files were converted to RNTuple while
preserving the source compression, with LZMA level 9 verified for both
containers. The TTree and RNTuple versions of each logical file were stored
side by side on the same CERN EOS instance and read by the same analysis.
The \ucache plugin handles RNTuple using the same two-tier structure as TTree. The byte
tier is format-blind, while the replica tier recognizes the container and
recompresses it using a structural rewriter that requires no ROOT libraries.
Replica coverage was 1453 of 1453 files in every run.

Two machines ran the full protocol: the workstation with 32 threads, a SATA
cache and a same-site source, and the desktop with 6 threads, an NVMe cache and
17.5\,ms latency to the source. Table~\ref{tab:rntuple} reports the wall-clock
times. Together, the four warm configurations provide the full comparison:
LZMA in both containers through the byte tier, and ZSTD-1 in both containers
through the replica tier. TTree with LZMA represents the current baseline.

\begin{table}[t]
\centering\small
\setlength{\tabcolsep}{4pt}
\caption{Performance comparison of the AGC workflow on 1453 files from \datasetL
  dataset in two containers (LZMA-9 in both), read by the same analysis from one
  EOS instance. Warm wall-clock times are measured with the page cache
  evicted. ``vs direct'' is the ratio to the direct-reading pass, which was run
  once per container and whose repeats varied by 7--8\% on the shared origin;
  ``Replica speedup'' is the ratio of warm-byte to warm-replica performance,
  comparing the two warm-cache tiers directly. The desktop reads over a 17.5\,ms
  wide-area link; the workstation reads over the local network.}
\label{tab:rntuple}
\begin{tabular}{llrrrrrr}
\toprule
 & & \multicolumn{3}{c}{wall (s)} & \multicolumn{2}{c}{vs direct} & Replica \\
\cmidrule(lr){3-5}\cmidrule(lr){6-7}
Machine & Container & direct & warm-byte & warm-replica & byte & replica & speedup \\
\midrule
\mW\ (32t, SATA) & RNTuple & 380 & 329 & 249 & 1.15 & 1.52 & 1.32 \\
\mW\ (32t, SATA) & TTree   & 864 & 548 & 337 & 1.58 & 2.56 & 1.63 \\
\mD\ (6t, NVMe)  & RNTuple & 2016 & 684 & 179 & 2.95 & 11.3 & 3.83 \\
\mD\ (6t, NVMe)  & TTree   & 3554 & 989 & 235 & 3.59 & 15.1 & 4.20 \\
\bottomrule
\end{tabular}
\end{table}

The results show a significant improvement from the RNTuple data format, which
reduces the gains provided by \ucache, especially for the byte cache when the
data source is nearby. This is consistent with the expected improvements in the
RNTuple storage layout. Nevertheless, \ucache remains effective for tightly
compressed data. On the workstation, the warm replica tier is 1.52 times faster
than direct reading for RNTuple and 2.56 times faster for TTree. On the desktop,
where every direct read crosses the wide-area link, both containers show much
larger gains: the byte cache is 3.0 to 3.6 times faster than direct reading, and
the replica tier is 11 to 15 times faster.

The difference between the two containers is explained primarily by the amount
of work performed rather than by additional stalling. On the workstation, the
warm byte-tier TTree pass executes $56.1\times10^{12}$ user-mode instructions,
compared with $33.8\times10^{12}$ for RNTuple. TTree therefore executes 1.66
times more instructions for identical physics. The two runs have nearly the
same IPC, 1.39 for TTree and 1.37 for RNTuple, which indicates that the
difference comes from the amount of work rather than from lower instruction
throughput. After recompression, the instruction counts fall to
$17.8\times10^{12}$ for TTree and $12.2\times10^{12}$ for RNTuple. The replica
tier therefore removes substantial decompression work from both formats, but
there is more work to remove from TTree.

The storage-layout difference is also visible at the system-call level. During
the cold fill, the RNTuple container is stored with 189 thousand write calls
averaging 588\,KiB, compared with 5.47 million calls averaging 25.4\,KiB for
TTree. The TTree fill therefore issues 29 times as many write calls for 1.25
times the bytes. The kernel merges writeback, so the difference observed by
the storage device is smaller than the system-call ratio. RDataFrame also opens
each RNTuple file once but each TTree file twice.

The two formats also produce different cache footprints after
recompression, and Table~\ref{tab:rntuple-footprint} separates the 3
effects behind the difference. The TTree byte cache carries a 1.19
page-rounding overhead (142.4\,GB cached for 119.6\,GB requested),
while RNTuple's large reads make rounding negligible. Recompressing dense
LZMA data to ZSTD-1 expands it in both containers, but by different
factors: the replica store is 1.04 times the byte cache for RNTuple and
1.22 times for TTree. Finally, the replicas supersede almost the entire
RNTuple byte cache, retaining 2.3\,GB, but a smaller fraction of the
TTree one, retaining 43.6\,GB. Together these give 120.6\,GB against
217.0\,GB for the same events.

\begin{table}[t]
\centering\small
\caption{Cache footprint for the two containers: the same 1453 files as
  Table~\ref{tab:rntuple}, default reclaim. The byte cache stores whole
  4\,KiB pages covering the requests; the replica store holds the
  recompressed data; the retained source bytes are the cached pages that
  replicas do not supersede.}
\label{tab:rntuple-footprint}
\begin{tabular}{lrr}
\toprule
 & RNTuple & TTree \\
\midrule
data requested by the analysis (GB) & 113.0 & 119.6 \\
byte cache after the cold pass (GB) & 113.7 & 142.4 \\
replica store (GB)                  & 118.4 & 173.5 \\
retained source bytes (GB)          & 2.3   & 43.6  \\
cache total after reclaim (GB)      & 120.6 & 217.0 \\
\bottomrule
\end{tabular}
\end{table}

Direct-reading performance also differs between the containers. The direct
TTree pass is 2.28 times slower than the direct RNTuple pass on the workstation
and 1.76 times slower on the desktop. This difference is specific to this
analysis and its configuration: RDataFrame opens each TTree file twice, and
Section~\ref{sec:rootscan} measures how strongly the TTree cost depends on
the ROOT version. The measurements also come from
non-interleaved runs against shared origins. We therefore treat this direct-read
difference as an observation rather than as evidence for data-conversion
decisions. None of the cache gains discussed above depend on that comparison.

The two containers also produce the same physics result. Across 124 histograms
and 3348 bins, the workstation campaign found zero entry-count mismatches.
Both containers sum to 320\,316\,108 histogram entries, and the largest relative
difference in weighted bin contents is $5\times10^{-12}$, consistent with
thread summation order.

The acceptance checks also exposed a small correctness issue in the first
campaign. On the first warm pass, the origin gate detected the RNTuple replica
tier fetching exactly 128\,KiB per file from the origin, corresponding to
0.17\% of served bytes. The cause was a reclaimed region inside the file
header. After fixing it, repeated measurements on both machines reported zero
origin bytes on every warm pass.

In conclusion, \ucache applies to both TTree and RNTuple without changing the
basic caching model. It fills within the direct-read performance target and
reaches full replica coverage for both containers. The magnitude of the gain,
however, depends on how much work remains after the container itself has
improved the access pattern. RNTuple already reduces storage operations and
executes fewer instructions for the same physics, so there is less work for the
cache to remove. Near the source, the byte tier improves performance by 1.58
times for TTree but only 1.15 times for RNTuple. Across the wide-area link, the
byte tier still provides gains of 3.0 to 3.6 times because both containers
benefit from avoiding repeated remote reads.

Compression is a separate effect. Recompressing cached data provides an
additional 1.3 to 3.8 times improvement within RNTuple and 1.6 to 4.2 times
within TTree. The limiting case is therefore an RNTuple dataset already stored
with a fast compression format and read from a nearby fast source. In that
case, little removable storage or decompression cost remains for a cache.
When the source is distant, or when the source uses dense compression for
archival efficiency, caching remains valuable. Analyst-side recompression can
then provide faster local decoding without requiring the shared archival copy
to use a less compact compression format. For the data available today, the
largest gains occur on the existing TTree datasets. And even for RNTuple
already stored in a fast codec, the least favorable configuration for a
cache, locality alone returned 1.6 and 3.5 times on the two machines
measured, and it helps wherever the machine can consume data faster than the
network path delivers, as the next subsection measures directly.

\subsubsection{RNTuple with a fast compression format}
\label{sec:rntuple-zstd}

To measure the limiting case directly, a third variant of the same files was
produced: the 1453-file TTree source converted to RNTuple with ZSTD level 1
(ROOT compression setting 501) instead of preserving LZMA-9, using the same
converter, the same writing ROOT version (6.36.02) and the same origin, so
the compression format is the only change. Every converted file was verified
against its source (entry counts and branch sums), and the whole set was
verified again reading it back from the origin. The ZSTD-1 set is 8.6\%
larger than the LZMA-9 RNTuple set and still 15\% smaller than the LZMA-9
TTree source. In this configuration neither of the cache's work-elimination
mechanisms can act, by construction: the source already uses the replica
tier's target compression format, so there is nothing to recompress, and the
byte cache preserves the container's own layout, which already reads in
large blocks ($\sim$2\,MiB requests to the origin, $\sim$440\,KiB reads
from the cache, and one file open per file). What remains to test is
locality alone.

Two machines ran this configuration against the same origin with the same
ROOT version: the cloud VM \mV, using all 24 cores with the cache on its
VM-local disk, and the CERN batch node \mFtwo, using 64 threads with the cache
on its NVMe system disk. The replica tier is absent on both by
construction. Table~\ref{tab:rntuple-zstd} shows the wall-clock results and
Table~\ref{tab:rntuple-zstd-warm} the anatomy of the warm passes.

\begin{table}[t]
\centering\small
\caption{The limiting-case measurement: the 1453-file ZSTD-1 RNTuple set
  read from the same CERN EOS origin on two machines, ROOT 6.36.02,
  wall-clock seconds of evicted passes. Both cold fills are inside the
  first-pass target, and both warm passes served zero bytes from the
  origin.}
\label{tab:rntuple-zstd}
\begin{tabular}{lrrrr}
\toprule
 & \multicolumn{2}{c}{\mV\ (24t)} & \multicolumn{2}{c}{\mFtwo\ (64t)} \\
\cmidrule(lr){2-3}\cmidrule(lr){4-5}
 & wall, s & vs direct & wall, s & vs direct \\
\midrule
direct, no cache & 176 & -- & 145 & -- \\
cold fill        & 184 & 1.04$\times$ & 137 & 0.94$\times$ \\
warm, byte tier  & 111 & 1.59$\times$ & \z42 & 3.50$\times$ \\
\bottomrule
\end{tabular}
\end{table}

\begin{table}[t]
\centering\small
\caption{Anatomy of the warm passes in Table~\ref{tab:rntuple-zstd}. The two
  machines serve an almost identical request stream (the same bytes in
  the same number of reads of the same size), which is what makes the
  comparison valid. The uncached (delivered) rate is derived from the cold pass's
  measured origin volume divided by the direct wall; the consumed rate is
  measured on the cache device during the warm pass.}
\label{tab:rntuple-zstd-warm}
\begin{tabular}{lrr}
\toprule
Warm byte-tier pass & \mV\ (24t) & \mFtwo\ (64t) \\
\midrule
data served from cache (GB)           & 119.7 & 119.7 \\
cache reads                           & 264\,070 & 266\,465 \\
mean read size (KiB)                  & 443 & 439 \\
delivered rate, uncached (MB/s)       & 679 & 824 \\
consumed rate, warm (MB/s)            & 1078 & 2882 \\
user-mode instructions ($10^{12}$)    & 12.4 & 6.9 \\
instructions per cycle                & 1.90 & 2.16 \\
gain from skipping page-cache eviction & 0.99$\times$ & 1.04$\times$ \\
\bottomrule
\end{tabular}
\end{table}

On the VM, the cache-filling pass costs 1.04 times the direct read, within
the first-pass target, and the warm byte-tier pass is 1.59 times faster
than direct reading with zero origin bytes. The mechanism is visible in the
rates: the uncached analysis consumed 679\,MB/s (the rate at which the
nearby source delivered data that hour), while the same analysis served
from the local cache consumed 1078\,MB/s. The analysis can consume data
faster than the network path delivered it, which is precisely the condition
under which a local cache layer is beneficial. The warm pass ends compute-bound:
12.4 trillion instructions at 1.90 instructions per cycle correspond to
roughly 89\% core occupancy, and eliminating page-cache eviction changes
nothing (0.99 times). Two independent readings therefore agree that the
cache has removed enough I/O that the processor, not the data path, limits
the run; no faster storage tier could improve it further.

On the batch node the same mechanism yields 3.50 rather than 1.59. Its
uncached pass consumed
824\,MB/s and its cold fill sustained 876\,MB/s from the origin (70\% of
the node's 10\,GbE interface, about the practical ceiling for bulk TCP at
standard frame size), so the direct pass was effectively limited by the
node's own network link rather than by the origin. Its warm pass consumed
2882\,MB/s from the local NVMe, with the page cache worth only a further
1.04 times: the node's own disk is 3.3 times faster than its share of the
network, which accounts for nearly all of the 3.50. The wider the gap between what the machine can consume and
what the network delivers, the more locality alone is worth. Part of that
gap is the processor itself: \mFtwo\ executes the same analysis in 6.9
trillion instructions against the VM's 12.4, at a higher rate per cycle.
The two machines differ in more than one way (a bare-metal Zen 5
against a virtualized Zen 4, with the hypervisor deciding which
instruction-set extensions the guest sees), so we do not attribute the
instruction gap to a single cause; what matters for the cache argument is
the consumption capability, not its origin.

Two bounds apply to the batch-node numbers. The runs are batch jobs on 64
of 192 hardware threads of a shared production node, so co-tenant load is
part of the measurement. And a ratio built on a delivery-limited direct
pass depends on what limited delivery. Here it was the interface, whose
ceiling is a property of the node, so 3.50 is close to the smallest gain
this machine can show against any remote source: no origin can deliver
past the link. The origins themselves delivered 0.31 to 0.88\,GB/s to
this cluster across the same night's campaigns, so against a slower
session the same machine would have measured more. The stable statement
is the measured gap itself: the cache served the warm pass at 2.9\,GB/s
while the network path delivered 0.8\,GB/s.

This measurement complements the LZMA-9 comparison above rather than
repeating it, and the two are bound by different resources. With LZMA-9
the warm pass is decompression-bound, so serving bytes locally gains little
near the source (1.15 times) and recompression is the productive
mechanism. With ZSTD-1, decompression is cheap, the uncached pass is bound
by data delivery, and locality alone returns 1.59 times on the cloud VM
and 3.50 times on the batch node: the cache's benefit tracks whichever
resource binds the uncached pass. Even in the configuration where the
container's geometry is close to ideal and recompression has nothing left
to remove (the least favorable case for a cache that this study can construct), the
cache still converts a delivery-bound analysis into a compute-bound one at
essentially no fill cost, and the faster the processor, the larger the
return.

\subsection{Sensitivity to the ROOT version}
\label{sec:rootscan}

Every measurement in this paper divides the analysis time between \ucache
and the software stack above it, and ROOT sets most of the analysis side. To
measure how much the results depend on the ROOT version, the TTree setup of
the previous subsection (the same 1453 files on the CERN-internal EOS
copy, the workstation, 32 threads, the same cache device, evicted passes)
was repeated on ROOT 6.34.04, 6.38.04 and 6.40.02, bracketing the 6.36
generation used on the Linux machines in this paper; the Mac mini campaigns
use 6.38.04, one of the versions scanned here. Each ROOT version is used as its
software distribution ships it, so the bundled XRootD client (5.7.3 to
6.0.3) and python version move with it. ROOT 6.40 links XRootD~6, which no
released \ucache version supports; that measurement used a development
build with preliminary XRootD~6 support. Table~\ref{tab:rootscan}
summarizes the series.

\begin{table}[t]
\centering\small
\caption{The TTree measurement of Section~\ref{sec:rntuple} repeated on 4
  ROOT versions: the same 1453 files, origin, machine (workstation, 32
  threads) and cache device; wall-clock seconds, evicted passes. Each ROOT
  version is used as its software bundle ships it, so the XRootD client
  version moves with ROOT; the 6.40.02 measurement used a development
  \ucache build with preliminary XRootD~6 support. The 6.36.02 column is
  the later of the two TTree campaigns of Section~\ref{sec:rntuple}; the
  earlier one, quoted there, measured direct 864, byte 548 and replica
  337\,s, agreement within the source-drift band. Every version ran with
  full replica coverage (1453 of 1453 files).}
\label{tab:rootscan}
\begin{tabular}{lrrrr}
\toprule
 & 6.34.04 & 6.36.02 & 6.38.04 & 6.40.02 \\
\midrule
XRootD client version & 5.7.3 & 5.8.3 & 5.9.1 & 6.0.3 \\
direct, no cache (s) & 1040 & 881 & 968 & 945 \\
warm, byte tier (s) & 723 & 530 & 619 & 611 \\
warm, replica tier (s) & 360 & 339 & 335 & 334 \\
byte-tier instructions ($10^{12}$) & 75.6 & 55.9 & 65.6 & 68.1 \\
byte-tier IPC & 1.36 & 1.42 & 1.42 & 1.47 \\
warm replica vs direct & 2.89$\times$ & 2.60$\times$ & 2.89$\times$ & 2.83$\times$ \\
\bottomrule
\end{tabular}
\end{table}

The ROOT version moves the analysis cost substantially. The warm byte-tier
pass spans 530 to 723\,s (a factor of 1.36) on identical data and
hardware. The instruction counts identify the difference as work rather
than execution efficiency: 55.9 to 75.6 trillion instructions at an IPC
that stays within 1.36 to 1.47. ROOT 6.36.02 is the fastest of the four:
6.34.04 executes 35\% more instructions for the same physics, and 6.38.04
and 6.40.02 give back part of the 6.36 improvement.

Attributing these differences to ROOT is justified by \ucache's own
counters, which show the same workload on every version: the warm byte-tier
pass makes 6.56 million cache reads on all four versions (the counts agree
within 160), the same number of vector-read chunks, the same 119.6\,GB
served, and exactly 2 file opens per file on every version, so the
double open is a property of the analysis framework across the whole
6.34--6.40 range rather than of one ROOT release.

The warm replica-tier pass is nearly indifferent to the ROOT version: it
stays within a 7.8\% band across the series, and within 1.6\% from 6.36
onward, while the direct pass moves by 18\% and the byte-tier pass by 36\%.
A slower ROOT leaves more removable work, so the cache's relative gain is
largest where ROOT is slowest: the warm replica pass is 2.83 to 2.89 times
faster than direct reading on 6.34, 6.38 and 6.40, against 2.56 to 2.60 on
6.36.02, and recompression is worth 2.01 times over the byte tier on
6.34.04 against 1.56 to 1.63 on 6.36.02. The TTree measurements elsewhere
in this paper use ROOT 6.36.02, the fastest of the four, so along this axis
the reported gains are conservative. Filling the cache stayed within the
first-pass design target on every version, with the cold pass at 1.02 to
1.09 times direct.

\section{Known limitations}
\label{sec:limits}

The cache does not improve every workflow or storage configuration. The cases
below describe where its benefits are limited, where a particular cache tier
becomes ineffective, and where the current implementation has known
constraints.

\begin{itemize}
\item \textbf{Single-pass workloads.} A job that reads a dataset only once
cannot benefit from faster repeated reads. It can still benefit from lower file-open
latency and from the automatic retry of failed reads, while adding only the small and bounded fill
overhead. The larger multiplicative speedups require the same data to be read
again. This is why iterative analysis development is the main target use case:
physicists naturally make repeated passes over the same data while developing and
refining an analysis.

\item \textbf{Compute-bound warm passes.} Improving the storage path cannot
reduce wall-clock time once the warm analysis is limited by computation,
whether in the analysis code or in decompression, rather than by I/O. On the
LZMA dataset, an internal change that reduced the byte tier's device
operations by a factor of 5.5 changed the wall-clock time by less than 1\%:
that pass was limited by LZMA decoding (Section~\ref{sec:codec}). In this regime, the storage system
is no longer the limiting resource. The plugin counters can identify this case
before time is spent tuning the cache or storage layout.

\item \textbf{The gain depends on the dataset.} The replica speedup varies
from 1.2--2.4 times on zlib-1 data to 1.6--4.7 times on LZMA-9 data
across the measured machines. The reason is that the replica tier removes decompression work, and the
amount of removable work depends on the compression format of the input data.
The cache itself does not change between these measurements. A single speedup
number is therefore not representative unless the dataset and its compression
format are also specified. The capacity arithmetic scales with the read
fraction the same way: these analyses read about 6\% of their datasets,
which is what lets a 1\,TB cache serve 10--20\,TB of remote data. An
analysis reading 30\% of its input needs a cache five times larger for the
same dataset; the speedup mechanisms are unchanged, only the sizing moves.

\item \textbf{Cache storage can be slower than the origin.} A cache helps only
if the local storage can serve the workload efficiently. We measured a
rate-limited cloud volume whose service class limited it at
$\sim$1000 read operations per second regardless of concurrency. On this
device, the byte tier is 6.9 times slower than no cache because its many
small reads reach the device operation-rate limit. The first-pass target of
$\le$1.1 times also cannot be met: filling the cache takes 6.1 times as
long as direct reading because writes are subject to the same operation-rate
limit.

The replica tier reduces the penalty with fewer and larger reads but does not
remove it, and its outcome on this volume is not even stable: 2 campaigns
3 weeks apart received throughput differing by 2.9 times on the
identical request pattern (Section~\ref{sec:geometry}). The bundled
device self-test (\texttt{ucache bench}) is intended to detect this condition
before deployment, and it applies equally to device classes not measured in
this paper (RAID arrays, network file systems, shared cluster storage)
before a cache is placed on them. On storage with such a low operation-rate
limit, the byte tier is not useful; the replica tier is required to obtain a
performance benefit.

\item \textbf{The working set can be larger than the cache.} We deliberately
oversubscribed the cache by $\sim$16\% on the 2.4\,TB dataset, twice: once
under the original eviction policy (release 0.18.1) and once under the
current one (0.21.0), which was changed because of what the first
measurement showed.

Under the original policy, least-recently-used eviction thrashed on the
linear scan: entries were evicted before the next pass needed them (177
during the cold fill, 532 per warm pass), so the cache never reached a
stable warm state and warm passes re-fetched a churning 20--41\% of the
bytes they served. Even so, \ucache remained 1.57 times faster than no
cache. In an analogous situation (the shipped \xcache default's
prefetch-driven demand exceeding its cache space by 1.68 times on the
zlib dataset), \xcache was 13.2 times slower than no cache
(Section~\ref{sec:xcache}). Replica recompression
behaved worst under this pressure: about 86\% of the replicas were evicted
before they were ever served, for a speedup of only 1.01 times.

Release 0.21.0 replaced that behavior: entries read within a protection
window (24\,h by default) are not eviction candidates, and a cache full of
protected entries declines new admissions instead of evicting the running
job's working set; declined files are relayed from the origin. Repeating
the same 1.16 oversubscription factor under this policy gives a partially
cached but \emph{stable} state: 1268 of 1456 files (87\%) cached, zero
evictions across the whole campaign, the same fixed 12.7\% of bytes relayed
from the origin on every warm pass, and 3 evicted warm passes agreeing
within 2.2\% at 1.86 times faster than no cache, with the fill still
inside the first-pass target ($0.99\times$) and the condition announced in
the client log and \texttt{ucache status}. The practical remedy is
unchanged: size the cache so that it can hold the working set, since the
relayed remainder still depends on origin latency and bandwidth on every
pass.
The failure mode of not doing so, however, is now a measured, graceful
degradation rather than thrashing.

\item \textbf{Concurrent analyses sharing one cache were not benchmarked.}
The cache is safe under concurrent use by construction (multiple
processes on one machine share it through atomic metadata updates, and the
crash-recovery and soak tests exercise exactly that), but every
performance measurement in this paper runs one analysis at a time. Two
users' working sets competing for one cache device and one eviction
budget is a realistic configuration this paper does not measure.

\item \textbf{Read-only inputs are assumed.} \ucache is a read cache for
immutable inputs: it never writes to the origin, and analysis outputs do
not pass through it. A freshness window (7 days by default) plus an
explicit revalidation mode handle the rare rewritten input file; workflows
that modify their inputs inside the analysis loop are outside the design
scope.

\item \textbf{One benchmark analysis.} Every measurement here runs the same
AGC $t\bar{t}$ analysis. The regime framework of Table~\ref{tab:regimes} is
how the results transfer to other workloads: an analysis with heavier
per-event computation is more compute-bound, so the cache matters less
(the warm ZSTD-1 passes of Section~\ref{sec:rntuple-zstd} show that end
state), while a lighter one is more delivery-bound, and the cache matters
more. The plugin's counters identify the regime for any workload. A second,
independent analysis workload from FCC studies is planned.


\item \textbf{Client support has a lower version bound.} The plugin engages on
XRootD~5 clients at version 5.6 or newer, and current builds target a glibc version corresponding to Enterprise Linux 9
(EL9) or newer. On older
clients, the plugin fails open to uncached reads. The analysis therefore
continues to run, but without caching.

\item \textbf{Origin independence depends on the reclaim setting.} With the
default reclaim, every quoted warm pass in this paper in which the cache held
the working set read zero bytes from the origin. Under \texttt{recompress\_reclaim = full}, warm passes
refetch the reads that replicas do not cover (about 12\,KB per file, or
0.01\% of served bytes, including the reads made when a file is opened),
so the space saving costs strict origin independence
(Section~\ref{sec:opscost}). All origin traffic is visible in the plugin
counters.
\end{itemize}

\section{Related work}
\label{sec:related}

\textbf{\xcache/\texttt{XrdPfc}}~\cite{bauerdick2014xcache,hanushevsky2019xcache}
is the established XRootD caching proxy. It is typically deployed and
administered as shared site infrastructure, whereas \ucache is intended for use
by an individual physicist on the analysis machine. Section~\ref{sec:xcache}
compares the two systems directly, including both the default and tuned
\xcache configurations. The measurements also suggest one design improvement
that could be useful in \texttt{XrdPfc} for sparse access patterns: decoupling the unit
of remote fetching from the unit retained in the cache. This would allow
batched sub-file reads while keeping cache residency bounded.

\textbf{\texttt{TTreeCache}}, provided by ROOT~\cite{brun1997root}, addresses a
different layer of the I/O path. It batches reads for a single process and
stores the prefetched data in memory. The \ucache plugin operates below \texttt{TTreeCache},
persists data across processes and days, and can serve any XRootD client.
\texttt{TTreeCache} remains enabled on top of \ucache throughout the measurements in
this paper, so the two mechanisms are complementary rather than alternatives.

\textbf{ROOT's local file cache.} ROOT can also cache remote files on local
disk without server-side support: \texttt{TFile::SetCacheFileDir} directs
every remotely opened file to be downloaded into a local directory and read
from there~\cite{brun1997root}. In deployment terms this is the closest
prior mechanism to \ucache, but the granularity differs: it transfers and
stores whole files, while the analyses measured here read about 6\% of the
bytes of the files they open. At that access density, whole-file caching
would move 1.8--2.4\,TB to local storage before these datasets are fully
cached (more than the 1\,TB cache devices used in this study), where
the page-granular byte cache stores 101--143\,GB, and every first pass over
a file transfers the whole file, regardless of how little of it the
analysis reads. The stored bytes also remain in the origin's compression
format, so there is no equivalent of the replica tier.

\textbf{Padulano et al.}~\cite{padulano2023daos,padulano2021caching}
developed an RNTuple-specific caching layer inside ROOT's I/O subsystem for an
Intel DAOS object store. Their system and \ucache share several ideas: both are
format-aware, cache only the data that are actually read, and retain those data
in compressed form. Their intended environments are otherwise different. The
DAOS cache resides on remote high-bandwidth cluster storage, is enabled
explicitly by the analysis, and targets RNTuple. The \ucache software is implemented as a
client plugin, operates transparently for XRootD clients, supports the existing
body of TTree data as well as RNTuple, and places the cache on the physicist's own
storage. It also includes cache lifecycle mechanisms such as eviction,
integrity validation and fail-open behavior, which were outside the scope of
the DAOS work.

\textbf{Tedeschi et al.}~\cite{tedeschi2024cpc} prototype distributed
RDataFrame analysis with \xcache staging for CMS. Their work targets caching
and data delivery at analysis-facility scale, in contrast to the end-user
caching model considered here.

\textbf{RNTuple}~\cite{blomer2020rntuple,lopezgomez2022rntuple} improves the
physical organization of ROOT data itself. Its storage layout reduces some of
the I/O costs that a cache can otherwise hide. Section~\ref{sec:rntuple}
measures how these format-level improvements change the gains provided by
\ucache and separates the effect of the container layout from the effect of
compression.

Finally, the use of \textbf{CMS Open Data as a computing benchmark} has
precedent both in the refereed literature~\cite{eschle2023julia} and in the AGC
effort~\cite{agc-repo,shadura2023agc}. The \textbf{subMIT facility
paper}~\cite{dalfonso2026submit} describes the analysis-facility environment
that motivates this type of cache. It also provides an independent performance
measurement against which we validate our experimental apparatus.

\section{Conclusions}
\label{sec:conclusions}

Physics analyses often read the same remote data repeatedly. As a result, each
pass may spend substantial time transferring and decompressing bytes that were
already transferred and decompressed in earlier passes. Our cache
implementation, \ucache, addresses this cost at the point where the analysis
runs. It is an XRootD client plugin that
stores only the pages actually read by the analysis on local storage. It can
also build replicas in which those data are recompressed and reordered for
faster repeated access. It requires no site infrastructure, no changes to
the analysis code, and no cooperation from the remote data source.

\begin{figure}[t]
\centering
\includegraphics[width=\textwidth]{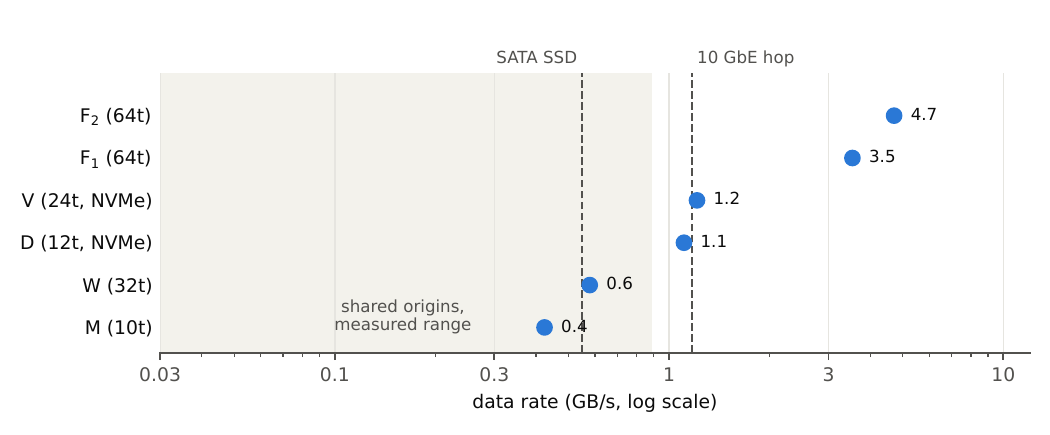}
\caption{What each machine consumes on its fastest measured warm pass (the
  replica-rate columns of Tables~\ref{tab:e2e-lzma} and~\ref{tab:e2e-zlib})
  against what the network layers deliver: the SATA interface ceiling
  measured on these devices, the measured 10\,GbE hop of
  Section~\ref{sec:xcache}, and the range the shared origins delivered
  across all campaigns (Sections~\ref{sec:origins}
  and~\ref{sec:rntuple-zstd}). Every machine except the Mac mini consumes
  at or above what a SATA device can serve, and the facility nodes consume
  several times what one 10\,GbE hop carries.}
\label{fig:consumption}
\end{figure}

The measurements support five main conclusions. The summary figures exclude
the deliberately rate-limited cloud volume, which was tested to expose the
operation-cost mechanism (Section~\ref{sec:geometry}) and is not a suitable
choice of cache storage. First, the byte cache can be created with very
little overhead. It fills during the first useful analysis pass; analysis
threads spend 0.1--1.2\% of that pass waiting on the fill machinery,
rising to 3.1\% only against the fastest source, which delivered data near
the cache device's sustained write rate, still well inside the 10\%
design target. The cache therefore does not require a separate staging step
before useful analysis can begin.

Second, repeated passes can be substantially faster and no longer depend on the
performance of the remote source. Serving cached source bytes is 1.6 to 8.8
times faster than direct remote reading in the measured configurations of
Tables~\ref{tab:e2e-lzma} and~\ref{tab:e2e-zlib}. Where
decompression is an important part of the warm-pass cost, replica serving adds
a further 1.2 to 4.7 times improvement. Warm-cache times are also much more
stable: they reproduce to about a percent, while direct-reading times fluctuate
with conditions at the remote source.

Third, the storage and construction costs are modest relative to the resulting
benefit. The cache stores only 6 to 9\% of the full dataset for these workloads.
On machines with processing capacity available during the first pass,
background replica construction overlaps with the fill, so reaching a fully
replicated cache costs about as much wall-clock time as the fill itself. The
additional work required for replicas is recovered within the first repeated
pass in the measured cases, except on a fully loaded host reading from a nearby
fast source, where background construction competes directly with the analysis.

Fourth, the size of the gain is determined by the workload and the available
storage and network resources. It is not tied to one ROOT container format or
one site. The same caching mechanism applies to both TTree and RNTuple. RNTuple
already removes some of the storage overhead present in TTree, so the additional
benefit from the byte cache is smaller when the data source is nearby and can
keep up with the analysis; when the machine consumes faster than the network
delivers, locality alone returned 3.5 times on a current-generation facility
node even for RNTuple already compressed with a fast format
(Section~\ref{sec:rntuple-zstd}). Recompression remains useful when
decompression is expensive, and caching remains valuable when the source is
distant. The largest gains in our
measurements occur for TTree, which remains the format of the large existing
data collections available for analysis.

Fifth, the measured consumption rates argue that the cache belongs on the
analysis machine itself, on storage faster than a SATA SSD
(Figure~\ref{fig:consumption}). Even the
desktop's consumer CPU saturates a SATA SSD and reaches the capacity of a
10\,GbE network path when running this analysis: with NVMe cache storage its warm replica
pass consumes 0.8 to 1.1\,GB/s at 6 to 12 threads, above the
$\sim$0.55\,GB/s a SATA device delivers and at the 1.17\,GB/s the
measured 10\,GbE link sustains (Section~\ref{sec:xcache}), and the
64-thread facility nodes consume 3 to 4.7\,GB/s. The two-machine experiment shows
the network case directly: with a fully warm cache one 10\,GbE hop away,
the warm pass took 245.5\,s against 163.0\,s for the local byte tier and
100.7\,s for the local replica tier. A 10\,GbE network is, in effect,
storage about twice as fast as the SATA interface this analysis saturates
at 12 threads, a margin the same machine's 0.96 to 1.1\,GB/s
consumption already covers. These rates are specific to this workload, and faster
network fabrics change the arithmetic; but at the measured per-core
consumption, a fast warm loop requires local cache storage.

The \ucache plugin does not replace shared site caching; it addresses a
different point in
the data path. On the same machine and storage device, its warm byte-serving
performance is similar to that of a tuned \xcache, while \ucache fills at
approximately direct-reading speed and uses about one third as much disk in the
measured workload. A shared \xcache can instead provide pooled storage capacity
and share cached data across many users. For an individual physicist with fast
local storage, however, \ucache provides a simple way to turn repeated remote
reads into local reads and, when recompression is beneficial, to reduce the
decompression cost as well. After the first pass, subsequent analysis passes
can therefore run primarily at the speed of the physicist's local storage and
compute resources rather than at the speed of the remote data source.

Viewed as a storage hierarchy, the analysis machine's NVMe sits between the
page cache and the site infrastructure, and \ucache is what makes it a
durable cache layer rather than scratch space. The measured rates order the
layers cleanly: the page cache is worth up to 2.4 times on a warm pass where
the working set fits in memory, the local NVMe devices deliver 1.1 to
6.5\,GB/s through \ucache's read paths at 0.05 to 0.06\,ms, one 10\,GbE
LAN hop delivers 1.17\,GB/s at 0.30\,ms, and the WAN origins delivered
0.03 to 0.5\,GB/s with large fluctuations. The local layer's advantage over
the network layers is structural rather than incidental: a shared network
path is provisioned for aggregate demand, so a single client's share of it
falls below a dedicated local device in any balanced deployment, and the
latency gap applies to every one of the millions of small reads a warm
analysis pass issues.

\section{Software and data availability}
\label{sec:availability}

The \ucache software is open source (MIT license)~\cite{ucache-repo}. The
measurements
span releases 0.18.1 through 0.21.0 (the XRootD~6 point of
Section~\ref{sec:rootscan} used a development build), mapped to
measurements in Appendix~\ref{app:repro}. The span does not weaken
comparisons: no ratio in this paper crosses releases (each divides
quantities measured in the same session with the same release), and
results taken with an earlier release are expected to be identical under
0.21.0 in the parts of the system they test; where the same measurement was
repeated across releases, the wall-clock times agreed to within 1\%
(Section~\ref{sec:origins}). The exception is eviction under an
overfilled cache, whose policy 0.21.0 deliberately changed;
Section~\ref{sec:limits} reports measurements of both policies.
Everything needed to repeat these measurements is public: the software,
the AGC RDataFrame analysis~\cite{agc-repo}, used unmodified, and the CMS
Open Data datasets, enumerated with DOIs in Appendix~\ref{app:datasets}.
Every run in this paper also recorded its release, configuration, and
acceptance counters in a machine-readable result, which is how the
per-run validity checks of Section~\ref{sec:protocol} are enforced.

\section*{Acknowledgments}

The authors thank the CMS Collaboration for the release of the datasets
used here, and CERN for the Open Data portal~\cite{cern-opendata} and the
\texttt{eospublic} service that serves it. Measurements at scale used the
subMIT analysis facility at MIT~\cite{dalfonso2026submit} and CERN
computing services: a workstation and cloud virtual machines, worker
nodes of the CMS global pool, and the EOS service that hosts the derived
RNTuple datasets. This work uses
exclusively public CMS Open Data (released under CC0); it has not been
reviewed or endorsed by the CMS Collaboration, and the authors bear sole
responsibility for its content. The authors thank the \xcache developers
for reviewing the \xcache comparison and its configuration
(Section~\ref{sec:xcache}). This work was supported by the
U.S.\ Department of Energy under Grant No.~DE-SC0011939.

\section*{Declaration on the use of AI-assisted technologies}

Claude Code (Anthropic's agentic coding tool) was used both in the
development of the \ucache software and in drafting this manuscript,
including its tables and figures, working from the authors' measurement
records and analysis notes. For the software, the authors specified the
requirements and architecture, directed each change, and validated the
implementation by its measured behavior: automated test suites
(unit, differential, crash-recovery, and fuzz testing under memory and race
detectors), integration tests against the real analysis frameworks and
storage services, physics validation showing that cached and direct
analyses produce identical results at full dataset scale, and the
measurement campaigns of this paper, in which every run records acceptance
criteria designed to expose incorrect behavior. The
authors reviewed, verified, and edited all manuscript content and take full
responsibility for it. No AI system is an author of this work.

\bibliographystyle{unsrtnat}
\bibliography{references}

\appendix

\section{Datasets}
\label{app:datasets}

\subsection{\datasetL: official CMS Open Data UL2016 NanoAODv9}

The \datasetL dataset maps each AGC sample role onto published UL2016
NanoAODv9 records (campaign
\texttt{RunIISummer20UL16NanoAODv9-106X\_mcRun2\_asymptotic\_v17-v1/v2},
NANOAODSIM), selected for generator/tune fidelity to the AGC samples; the
mapping and its physics caveats are documented with the benchmark kit.
This is an I/O study: the dataset is a faithful I/O replacement, not a
physics reproduction of the AGC result. One dataset
(\texttt{WJetsToLNu\_1J}, record 69717) is only partially available on the
portal (30 of an advertised 117 files; the record page's public file
count describes the full CMS dataset, while the portal's internal
availability field and the file index carry the true figure); the dataset
uses the 30 published files, keeping every input reproducible from the
portal alone. Totals as used: \textbf{1456 files, 2412\,GB,
$1.30\times10^9$ events} across 26 records.
Table~\ref{tab:datasets-ul2016} lists every record with its DOI.

\begin{table}[p]
\centering\scriptsize
\setlength{\tabcolsep}{3pt}
\caption{The \datasetL dataset: 26 CMS Open Data UL2016 NanoAODv9 records.
All datasets are campaign
\texttt{RunIISummer20UL16NanoAODv9-}\allowbreak\texttt{106X\_mcRun2\_asymptotic\_v17-v1/NANOAODSIM}
($^{\dagger}$: \texttt{-v2}). Files/GiB are as used; for record 69717 that
is the 30 published files (45.5\,M events measured from the files), not the
record page's advertised count.}
\label{tab:datasets-ul2016}
\begin{tabular}{>{\raggedright\arraybackslash}p{2.0cm}>{\raggedright\arraybackslash}p{4.6cm}rrrl}
\toprule
role & dataset & rec.\ & files & GiB & DOI \\
\midrule
ttbar/nominal & \texttt{TTToSemiLeptonic\_\allowbreak TuneCP5\_\allowbreak 13TeV-\allowbreak powheg-\allowbreak pythia8} & 67993 & 138 & 287.1 & \texttt{10.7483/OPENDATA.CMS.4J3Y.1CME} \\
 & \texttt{TTToHadronic\_\allowbreak TuneCP5\_\allowbreak 13TeV-\allowbreak powheg-\allowbreak pythia8} & 67841 & 146 & 211.6 & \texttt{10.7483/OPENDATA.CMS.PW7L.V890} \\
 & \texttt{TTTo2L2Nu\_\allowbreak TuneCP5\_\allowbreak 13TeV-\allowbreak powheg-\allowbreak pythia8} & 67801 & 49 & 84.9 & \texttt{10.7483/OPENDATA.CMS.4RTG.JPI2} \\
ttbar/scaledown & \texttt{TTToHadronic\_\allowbreak TuneCP5down\_\allowbreak 13TeV-\allowbreak powheg-\allowbreak pythia8} & 67847 & 48 & 76.7 & \texttt{10.7483/OPENDATA.CMS.JKOG.DRBO} \\
 & \texttt{TTToSemiLeptonic\_\allowbreak TuneCP5down\_\allowbreak 13TeV-\allowbreak powheg-\allowbreak pythia8} & 68001 & 21 & 38.5 & \texttt{10.7483/OPENDATA.CMS.1MRK.EV4Z} \\
 & \texttt{TTTo2L2Nu\_\allowbreak TuneCP5down\_\allowbreak 13TeV-\allowbreak powheg-\allowbreak pythia8} & 67807 & 19 & 35.6 & \texttt{10.7483/OPENDATA.CMS.IVG0.LAWQ} \\
 & \texttt{TTToSemiLeptonic\_\allowbreak hdampDOWN\_\allowbreak TuneCP5\_\allowbreak 13TeV-\allowbreak powheg-\allowbreak pythia8} & 67977 & 51 & 119.8 & \texttt{10.7483/OPENDATA.CMS.20N3.6DSY} \\
 & \texttt{TTToHadronic\_\allowbreak hdampDOWN\_\allowbreak TuneCP5\_\allowbreak 13TeV-\allowbreak powheg-\allowbreak pythia8} & 67825 & 88 & 80.4 & \texttt{10.7483/OPENDATA.CMS.8E9T.I780} \\
 & \texttt{TTTo2L2Nu\_\allowbreak hdampDOWN\_\allowbreak TuneCP5\_\allowbreak 13TeV-\allowbreak powheg-\allowbreak pythia8} & 67775 & 18 & 35.1 & \texttt{10.7483/OPENDATA.CMS.EGFN.XVRA} \\
ttbar/scaleup & \texttt{TTToSemiLeptonic\_\allowbreak TuneCP5up\_\allowbreak 13TeV-\allowbreak powheg-\allowbreak pythia8} & 68005 & 56 & 115.1 & \texttt{10.7483/OPENDATA.CMS.1BOB.74YG} \\
 & \texttt{TTToHadronic\_\allowbreak TuneCP5up\_\allowbreak 13TeV-\allowbreak powheg-\allowbreak pythia8} & 67851 & 41 & 82.4 & \texttt{10.7483/OPENDATA.CMS.FKTA.DBXG} \\
 & \texttt{TTTo2L2Nu\_\allowbreak TuneCP5up\_\allowbreak 13TeV-\allowbreak powheg-\allowbreak pythia8} & 67811 & 42 & 35.3 & \texttt{10.7483/OPENDATA.CMS.SSSP.MTOV} \\
 & \texttt{TTToSemiLeptonic\_\allowbreak hdampUP\_\allowbreak TuneCP5\_\allowbreak 13TeV-\allowbreak powheg-\allowbreak pythia8} & 67979 & 58 & 121.0 & \texttt{10.7483/OPENDATA.CMS.1I4V.QODF} \\
 & \texttt{TTToHadronic\_\allowbreak hdampUP\_\allowbreak TuneCP5\_\allowbreak 13TeV-\allowbreak powheg-\allowbreak pythia8} & 67827 & 44 & 83.0 & \texttt{10.7483/OPENDATA.CMS.6F7W.GLPN} \\
 & \texttt{TTTo2L2Nu\_\allowbreak hdampUP\_\allowbreak TuneCP5\_\allowbreak 13TeV-\allowbreak powheg-\allowbreak pythia8} & 67777 & 24 & 36.8 & \texttt{10.7483/OPENDATA.CMS.C2PE.1351} \\
ttbar/ME\_\allowbreak var & \texttt{TTJets\_\allowbreak TuneCP5\_\allowbreak 13TeV-\allowbreak amcatnloFXFX-\allowbreak pythia8} & 67731 & 109 & 186.8 & \texttt{10.7483/OPENDATA.CMS.QHPN.KEAC} \\
ttbar/PS\_\allowbreak var & \texttt{TT\_\allowbreak TuneCH3\_\allowbreak 13TeV-\allowbreak powheg-\allowbreak herwig7} & 68047 & 104 & 140.7 & \texttt{10.7483/OPENDATA.CMS.TVS5.3SG0} \\
wjets/nominal & \texttt{WJetsToLNu\_\allowbreak 1J\_\allowbreak TuneCP5\_\allowbreak 13TeV-\allowbreak amcatnloFXFX-\allowbreak pythia8} & 69717 & 30 & 49.7 & \texttt{10.7483/OPENDATA.CMS.I4BK.06EU} \\
 & \texttt{WJetsToLNu\_\allowbreak 0J\_\allowbreak TuneCP5\_\allowbreak 13TeV-\allowbreak amcatnloFXFX-\allowbreak pythia8} & 69715 & 97 & 141.1 & \texttt{10.7483/OPENDATA.CMS.EYSK.TSR8} \\
 & \texttt{WJetsToLNu\_\allowbreak 2J\_\allowbreak TuneCP5\_\allowbreak 13TeV-\allowbreak amcatnloFXFX-\allowbreak pythia8} & 69719 & 63 & 114.7 & \texttt{10.7483/OPENDATA.CMS.FZFI.0QD7} \\
single\_\allowbreak top\_\allowbreak s\_\allowbreak chan/nominal & \texttt{ST\_\allowbreak s-\allowbreak channel\_\allowbreak 4f\_\allowbreak leptonDecays\_\allowbreak TuneCP5\_\allowbreak 13TeV-\allowbreak amcatnlo-\allowbreak pythia8} & 64635 & 19 & 8.1 & \texttt{10.7483/OPENDATA.CMS.Y5N3.PYKM} \\
 & \texttt{ST\_\allowbreak s-\allowbreak channel\_\allowbreak 4f\_\allowbreak hadronicDecays\_\allowbreak TuneCP5\_\allowbreak 13TeV-\allowbreak amcatnlo-\allowbreak pythia8} & 64633 & 23 & 7.8 & \texttt{10.7483/OPENDATA.CMS.FXUV.43I0} \\
single\_\allowbreak top\_\allowbreak t\_\allowbreak chan/nominal & \texttt{ST\_\allowbreak t-\allowbreak channel\_\allowbreak top\_\allowbreak 4f\_\allowbreak InclusiveDecays\_\allowbreak TuneCP5\_\allowbreak 13TeV-\allowbreak powheg-\allowbreak madspin-\allowbreak pythia8} & 64759 & 86 & 97.4 & \texttt{10.7483/OPENDATA.CMS.448S.OXEJ} \\
 & \texttt{ST\_\allowbreak t-\allowbreak channel\_\allowbreak antitop\_\allowbreak 4f\_\allowbreak InclusiveDecays\_\allowbreak TuneCP5\_\allowbreak 13TeV-\allowbreak powheg-\allowbreak madspin-\allowbreak pythia8} & 64659 & 35 & 47.4 & \texttt{10.7483/OPENDATA.CMS.D885.EHVZ} \\
single\_\allowbreak top\_\allowbreak tW/nominal & \texttt{ST\_\allowbreak tW\_\allowbreak antitop\_\allowbreak 5f\_\allowbreak inclusiveDecays\_\allowbreak TuneCP5\_\allowbreak 13TeV-\allowbreak powheg-\allowbreak pythia8\,$^{\dagger}$} & 64825 & 23 & 4.6 & \texttt{10.7483/OPENDATA.CMS.GOSF.G1IA} \\
 & \texttt{ST\_\allowbreak tW\_\allowbreak top\_\allowbreak 5f\_\allowbreak inclusiveDecays\_\allowbreak TuneCP5\_\allowbreak 13TeV-\allowbreak powheg-\allowbreak pythia8\,$^{\dagger}$} & 64881 & 24 & 4.5 & \texttt{10.7483/OPENDATA.CMS.B9V5.WXFJ} \\
\midrule
\multicolumn{3}{l}{\textbf{total}} & \textbf{1456} & \textbf{2246.1} & \\
\bottomrule
\end{tabular}
\end{table}

For the Fermilab origin runs the same records were replicated to
\texttt{T1\_US\_FNAL\_Disk} with standard Rucio rules (the Open Data copy
is itself a Rucio-managed replica of the same logical file names (LFNs), so site doors serve
identical files, verified per-file by size before any benchmark, with
4 tape-resident files at FNAL excluded: 0.27\% of files, 0.22\% of
bytes).

\subsection{\datasetZ: AGC 2015 NanoAOD}

The \datasetZ dataset is the AGC project's fixed input list~\cite{agc-repo}:
787 NanoAOD files (1777\,GB, $0.94\times10^9$ events) derived by the AGC
team from 13 published CMS 2015 Open Data MINIAODSIM records and served
from the CERN Open Data EOS instance. The derived files are not themselves
a portal record; the parent records and their DOIs are listed in
Table~\ref{tab:datasets-agc}, and the exact file list is fixed by the AGC
repository's \texttt{nanoaod\_inputs.json}.

\begin{table}[p]
\centering\scriptsize
\setlength{\tabcolsep}{3pt}
\caption{Parent records of the \datasetZ dataset: 13 CMS 2015 Open Data
MINIAODSIM datasets (campaign
\texttt{RunIIFall15MiniAODv2-}\allowbreak\texttt{PU25nsData2015v1\_76X\_mcRun2\_asymptotic\_v12*/MINIAODSIM}),
from which the AGC team derived the 787-file NanoAOD input list. ``AGC
files/events'' are the derived dataset's share.}
\label{tab:datasets-agc}
\begin{tabular}{>{\raggedright\arraybackslash}p{2.0cm}>{\raggedright\arraybackslash}p{4.2cm}rrrl}
\toprule
sample & parent dataset & rec.\ & AGC files & AGC events & DOI \\
\midrule
ttbar/nominal & \texttt{TT\_\allowbreak TuneCUETP8M1\_\allowbreak 13TeV-\allowbreak powheg-\allowbreak pythia8} & 19980 & 84 & 94\,555\,484 & \texttt{10.7483/OPENDATA.CMS.JJEM.1DKC} \\
 & \texttt{TT\_\allowbreak TuneCUETP8M1\_\allowbreak 13TeV-\allowbreak powheg-\allowbreak pythia8} & 19981 & 159 & 181\,523\,643 & \texttt{10.7483/OPENDATA.CMS.4BI9.J6N1} \\
ttbar/scaledown & \texttt{TT\_\allowbreak TuneCUETP8M1\_\allowbreak 13TeV-\allowbreak powheg-\allowbreak scaledown-\allowbreak pythia8} & 19983 & 32 & 39\,329\,663 & \texttt{10.7483/OPENDATA.CMS.5FQF.EOZ1} \\
ttbar/scaleup & \texttt{TT\_\allowbreak TuneCUETP8M1\_\allowbreak 13TeV-\allowbreak powheg-\allowbreak scaleup-\allowbreak pythia8} & 19985 & 33 & 38\,424\,467 & \texttt{10.7483/OPENDATA.CMS.8NKT.I5JF} \\
ttbar/ME\_\allowbreak var & \texttt{TT\_\allowbreak TuneCUETP8M1\_\allowbreak 13TeV-\allowbreak amcatnlo-\allowbreak pythia8} & 19978 & 16 & 19\,098\,219 & \texttt{10.7483/OPENDATA.CMS.HRUJ.JHRZ} \\
ttbar/PS\_\allowbreak var & \texttt{TT\_\allowbreak TuneEE5C\_\allowbreak 13TeV-\allowbreak powheg-\allowbreak herwigpp} & 19999 & 15 & 19\,337\,064 & \texttt{10.7483/OPENDATA.CMS.KGUZ.PIUI} \\
single\_\allowbreak top\_\allowbreak s\_\allowbreak chan/nominal & \texttt{ST\_\allowbreak s-\allowbreak channel\_\allowbreak 4f\_\allowbreak InclusiveDecays\_\allowbreak 13TeV-\allowbreak amcatnlo-\allowbreak pythia8} & 19394 & 6 & 2\,867\,199 & \texttt{10.7483/OPENDATA.CMS.V11Z.ZIBW} \\
single\_\allowbreak top\_\allowbreak t\_\allowbreak chan/nominal & \texttt{ST\_\allowbreak t-\allowbreak channel\_\allowbreak top\_\allowbreak 4f\_\allowbreak inclusiveDecays\_\allowbreak 13TeV-\allowbreak powhegV2-\allowbreak madspin-\allowbreak pythia8\_\allowbreak TuneCUETP8M1} & 19408 & 58 & 70\,373\,744 & \texttt{10.7483/OPENDATA.CMS.R48B.H30Q} \\
 & \texttt{ST\_\allowbreak t-\allowbreak channel\_\allowbreak antitop\_\allowbreak 4f\_\allowbreak inclusiveDecays\_\allowbreak 13TeV-\allowbreak powhegV2-\allowbreak madspin-\allowbreak pythia8\_\allowbreak TuneCUETP8M1} & 19406 & 32 & 38\,932\,192 & \texttt{10.7483/OPENDATA.CMS.7CS4.MIKT} \\
single\_\allowbreak top\_\allowbreak tW/nominal & \texttt{ST\_\allowbreak tW\_\allowbreak antitop\_\allowbreak 5f\_\allowbreak inclusiveDecays\_\allowbreak 13TeV-\allowbreak powheg-\allowbreak pythia8\_\allowbreak TuneCUETP8M1} & 19412 & 2 & 999\,400 & \texttt{10.7483/OPENDATA.CMS.JG3K.U1HH} \\
 & \texttt{ST\_\allowbreak tW\_\allowbreak top\_\allowbreak 5f\_\allowbreak inclusiveDecays\_\allowbreak 13TeV-\allowbreak powheg-\allowbreak pythia8\_\allowbreak TuneCUETP8M1} & 19419 & 1 & 1\,000\,000 & \texttt{10.7483/OPENDATA.CMS.EKZ2.EXSU} \\
wjets/nominal & \texttt{WJetsToLNu\_\allowbreak TuneCUETP8M1\_\allowbreak 13TeV-\allowbreak amcatnloFXFX-\allowbreak pythia8} & 20547 & 191 & 237\,908\,687 & \texttt{10.7483/OPENDATA.CMS.YRJY.OK2Z} \\
 & \texttt{WJetsToLNu\_\allowbreak TuneCUETP8M1\_\allowbreak 13TeV-\allowbreak amcatnloFXFX-\allowbreak pythia8} & 20548 & 158 & 195\,810\,412 & \texttt{10.7483/OPENDATA.CMS.FLGU.07DD} \\
\midrule
\multicolumn{3}{l}{\textbf{total}} & \textbf{787} & \textbf{940\,160\,174} & \\
\bottomrule
\end{tabular}
\end{table}

\section{Software versions and reproduction}
\label{app:repro}

\begin{itemize}
\item \ucache (\texttt{xrd-ucache})~\cite{ucache-repo}, by measurement:
the end-to-end, mechanism and per-machine results
(Sections~\ref{sec:endtoend}--\ref{sec:geometry}) use release 0.19.1,
except the Mac mini campaigns (release 0.20.0) and the \mV\ row of
Table~\ref{tab:e2e-lzma} (release 0.21.0); the open-latency
measurement and the \xcache comparison use release 0.18.1; the storage
self-test records of Table~\ref{tab:bench} use the release current at
each machine's characterization (0.18.3--0.19.1, identical I/O paths
across them); the RNTuple comparison (Section~\ref{sec:rntuple}) uses
releases 0.18.1 through 0.19.1; the ROOT~6.40 point of
Section~\ref{sec:rootscan} used a development build with preliminary
XRootD~6 support; and the over-run measurement of
Section~\ref{sec:limits} deliberately pairs 0.18.1 and 0.21.0, the
eviction-policy change between them being its subject. Each table states
the release of its rows, and each
run's machine-readable result records the version that produced it.
\item ROOT 6.36.02 (LCG\_108), XRootD client 5.8.3, on EL9-class Linux
on the workstation, both facility machines and the cloud VM; the desktop, also
EL9-class, runs its own ROOT build of the 6.36 generation (reading
RNTuple requires 6.36 or newer); the Mac mini runs ROOT 6.38.04 and
XRootD client 5.9.1 from MacPorts on macOS/arm64. The subMIT cross-check
used the paper's own ROOT 6.32.08~\cite{dalfonso2026submit}.
\item Analysis: AGC \texttt{cms-open-data-ttbar} RDataFrame
implementation~\cite{agc-repo}; input lists as in
Appendix~\ref{app:datasets}.
\item \xcache comparison (Section~\ref{sec:xcache}): XRootD 5.9.6
\texttt{XrdPfc}; shipped defaults verified in source and in the server's
configuration dump (128\,KiB blocks, \texttt{prefetch 10},
\texttt{pfc.ram 1g}); tuned configuration \texttt{prefetch 0},
\texttt{pfc.ram 8g}, \texttt{oss.preread 32 limit 16m qsize 256}.
Two-machine runs: 24-core EPYC client (91\,GB RAM, NVMe cache disk),
cache host one 10\,GbE/0.3\,ms hop away; rate-limited-volume runs on the
same client with only the cache device changed.
\item Protocol per Section~\ref{sec:protocol}: page-cache eviction
verified with \texttt{mincore} before every timed warm pass (except on \mM,
per Section~\ref{sec:protocol}); where a campaign repeated its warm passes,
the median of the 3 evicted passes is quoted; validity checks (origin
share, CRC, fail-open, validation, coverage) recorded per run in each
run's machine-readable result.
\end{itemize}

\end{document}